\documentclass[a4paper,11pt]{article}

\usepackage{imakeidx}
\makeindex[columns=2, title=Alphabetical Index, intoc]
\usepackage{jheppub} % for details on the use of the package, please see the JINST-author-manual

\usepackage{float}

\usepackage{amsmath}
\usepackage{physics}
\usepackage{amssymb}
\usepackage{slashed}

\usepackage{array}

\DeclareUnicodeCharacter{27E8}{\langle}
\DeclareUnicodeCharacter{27E9}{\rangle}

\usepackage{tikz-feynman}
\usepackage{mathdots}

\usepackage{cleveref}
\crefname{equation}{Eq}{Eqs}
\crefrangelabelformat{equation}{(#3#1#4--#5#2#6)}
\crefname{table}{Table}{Tables}
\crefrangelabelformat{table}{#3#1#4--#5#2#6}

\indexsetup{firstpagestyle=indexpagestyle}

\def\UoE{Higgs Centre for Theoretical Physics, University of Edinburgh, Edinburgh, EH9 3FD, UK}
\def\KIT{Institute for Theoretical Particle Physics, KIT, Wolfgang-Gaede-Straße 1, 76131, Karlsruhe, Germany}

\preprint{
\begin{flushright}
P3H-26-014,
TTP26-004
\end{flushright}
}

\let\oldsqrt\sqrt
\def\sqrt{\mathpalette\DHLhksqrt}
\def\DHLhksqrt#1#2{%
\setbox0=\hbox{$#1\oldsqrt{#2\,}$}\dimen0=\ht0
\advance\dimen0-0.3\ht0
\setbox2=\hbox{\vrule height\ht0 depth -\dimen0}%
{\box0\lower0.4pt\box2}}

\newcommand{\be}{\begin{equation}}
\newcommand{\ee}{\end{equation}}
\newcommand{\ep}{\epsilon}
\newcommand{\m}{\phantom{-}}

\title{Polarization structure and spin covariance of massive vector-boson amplitudes in QCD}

\author[a]{Giuseppe De Laurentis,}
\author[b]{Kirill Melnikov,}
\author[b]{Matteo Tresoldi}

\emailAdd{giuseppe.delaurentis@ed.ac.uk}
\emailAdd{kirill.melnikov@kit.edu}
\emailAdd{matteo.tresoldi@partner.kit.edu}
\affiliation[a]{\UoE}
\affiliation[b]{\KIT}

\abstract{
Nearly thirty years ago, Bern, Dixon and Kosower computed all helicity amplitudes for the annihilation of an electron–positron pair into four QCD partons through an electroweak vector boson. More recently, the leading-color two-loop amplitudes for the same process were obtained. 
When such amplitudes are expressed in the massless spinor-helicity formalism, they effectively correspond to the decay of a \emph{transversely polarized} vector boson. However, for several reasons, it is highly desirable to extend these calculations to the case where the polarization of the vector boson is \emph{longitudinal}.
Due to the complexity of such computations, repeating them to obtain the result for the ``missing'' polarization of the electroweak boson is a significant undertaking even at one loop. Besides, when attempting new higher-loop computations, it is beneficial to identify the minimal set of quantities (e.g.~form factors) that must be determined to obtain the full amount of physically-relevant information. 
In this paper, we show that amplitudes involving vector-boson decays to massless leptons---although they appear to project onto the transverse polarization---still encode the full information about all polarization states of the vector boson, including the longitudinal one.
This follows from the little-group (spin) covariance of the amplitude, which allows us---largely through simple replacement rules---to rewrite the helicity amplitudes as a matrix with open $\mathrm{SU}(2)$ spin indices in the massive spinor-helicity (or spin-spinor) formalism. Therefore, knowledge of an amplitude for any polarization component suffices to reconstruct the full covariant matrix. 
}

\keywords{QCD corrections}

\allowdisplaybreaks
\begin{document}
\maketitle
% \flushbottom  % This causes vertical spaces between paragraphs

%-----------------------------------------------------
%                     FIGURES
%-----------------------------------------------------
\newcommand{\FigOneLoopJdiag}[2]{
    % #1 = position of the image
    % #2 = Caption
    \begin{figure}[#1]
        \centering
        \begin{tikzpicture}
            \begin{feynman}
                % Blobs definitions
                \vertex [blob, fill=none] (c) {1L};
                \vertex [blob, right=3.6cm of c] (fc) {};
                \diagram*{
                    (c) -- [boson, edge label=$q$] (fc);
                };
                % --- Initial-state legs ---
                \foreach \angle in {150,170,190,210}
                {
                    \vertex at ($(c)+(\angle:1.5cm)$) (i\angle);
                    \diagram*{
                        (i\angle) -- [plain] (c);
                    };
                }
                % --- Final-state legs ---
                \foreach \angle in {320,340,0,20,40}
                {
                    \vertex at ($(fc)+(\angle:1.5cm)$) (f\angle);
                    \diagram*{
                        (fc) -- [plain] (f\angle);
                    };
                }
                % --- Brace 1 cm below the diagram ---
                \draw[decorate, decoration={brace, mirror, amplitude=0.2cm}]
                    ($(i170 |- c) + (0,-1cm)$) -- ($(c) + (0.9cm,-1cm)$)
                    node[midway, below=0.25cm] {$A^{\mu}_{V\to 4j }$};
                \draw[decorate,decoration={brace, mirror, amplitude=0.2cm}]
                    ($(c) + (1cm,-1cm)$) -- ($(fc) + (-1cm,-1cm)$)
                    node[midway, below=0.25cm] {$D_V(q) \, \rho_{\mu\nu}$};
                \draw[decorate, decoration={brace, mirror, amplitude=0.2cm}]
                    ($(fc) + (-0.9cm,-1cm)$) -- ($(f0 |- fc) + (0,-1cm)$)
                    node[midway, below=0.25cm] {$J^\nu(q)$};
                % --- Vertical dashed lines ---
                \draw[dashed]
                    ($ (c |- f40) + (0.95cm,0.25cm) $) -- ($ (c |- f320) + (0.95cm,-0.25cm) $);
                \draw[dashed]
                    ($ (fc |- f40) - (0.95cm,-0.25cm) $) -- ($ (fc |- f320) - (0.95cm,0.25cm) $);
            \end{feynman}
        \end{tikzpicture}
        %=========================
        % CAPTION
        %=========================
        \caption{#2}
    \end{figure}
}

%-----------------------------------------------------
%                     SECTIONS
%-----------------------------------------------------
%%%%%%%%%%%%%%%%%%%%%%%%%%%%%%%%%%%%%%%%%%%%%%%%%%%%%%%%%%%%%%
%                        INTRODUCTION                        %
%%%%%%%%%%%%%%%%%%%%%%%%%%%%%%%%%%%%%%%%%%%%%%%%%%%%%%%%%%%%%%

\section{Introduction}

Loop amplitudes for transitions of electroweak vector bosons to QCD partons play an important role in particle physics. They are used to describe processes that involve production and decays of massive electroweak gauge bosons and (off-shell) photons at lepton and hadron colliders. They are also used for computing cross sections of more complex processes, such as the associated production of the Higgs boson and a vector boson at the LHC, and the process of the Higgs boson production in weak boson fusion.  

Many such amplitudes were computed at different orders in QCD perturbation theory in the past. A special place in this effort is occupied by Ref.~\cite{Bern_1998} where the analytic computation of the one-loop amplitudes for $e^+e^- \to 4~{\rm jets}$ was performed. The analytic results reported in this paper are very compact and, as discussed in Ref.~\cite{Bern_1998}, achieving such a degree of simplicity and compactness of the final result required  significant effort that in many ways anticipated later developments in the field of scattering amplitudes. Only recently, the two-loop leading-color amplitudes were obtained in Ref.~\cite{Abreu:2021asb} in terms of three form factors and later expressed in spinor-helicity variables in Ref.~\cite{DeLaurentis:2025dxw}, providing compact analytic expressions in a form analogous to the one-loop results of Ref.~\cite{Bern_1998}.

In fact, even today, nearly thirty years later,  repeating the calculation of Ref.~\cite{Bern_1998} is a challenging enterprise. Indeed, if standard computational tools focused on rewriting the contributing Feynman diagrams in terms of independent scalar (master) integrals are used, intermediate results grow very rapidly and become increasingly  difficult to manage, despite all the progress with the one-loop computations in the past thirty years \cite{Buccioni:2019sur,Denner:2016kdg,Berger:2008sj,Ossola:2007ax,Ossola:2006us,Alwall_2011,Alwall_2014,Actis_2013,Actis_2017,Cullen_2012,Cullen_2014,OL2015,Buccioni_2018,Cascioli_2012,Berger_2008,Ellis_2012,Peraro_2014,Mastrolia_2010,Borowka_2018,Forde_2007,Bern_1994_1,Bern_2011,Hahn_2000,Badger_2009,Badger_2008,Ellis_2008,Giele_2008,Hameren_2009,Bevilacqua_2009,Bevilacqua_2013,Binoth_2008uq,Giele_2009ui,Giele_2008i}.

One-loop amplitudes of the type computed in Ref.~\cite{Bern_1998} are  needed to calculate NNLO QCD corrections to various physical processes. For example, they are required for studies of the associated production of the Higgs boson and a vector boson together with a jet $pp \to HV+j$, or for studies of the production of the Higgs boson in weak boson fusion in association with an extra jet $pp \to H+3j$, or to describe the  production of a vector boson in association with jets, followed by the decay of the vector boson into heavy fermions, e.g. $pp \to 2j+(V \to \tau^+ \tau^-)$. Furthermore, 
they are important ingredients for studying  polarization effects in  vector boson production  at the LHC. 

However, for all these cases, the results obtained in Refs~\cite{Bern_1998} are insufficient. This is related to the fact that the matrix element of the vector current with respect to massless electron and positron spinors $[6|\gamma^\mu|5\rangle$ was used in Refs~\cite{Bern_1998,DeLaurentis:2025dxw} as a proxy for the %transverse 
polarization vector of the decaying (off-shell) electroweak vector boson. While this procedure correctly captures amplitudes for the process $e^+e^- \to 4 j$, for more complex processes additional amplitudes are required. Indeed, considering for definiteness the case of the Higgs boson production in weak boson fusion, it is easy to realize that the polarization of the $t$-channel off-shell vector boson cannot be described in the same way. One may say, colloquially, that the longitudinal polarization of the intermediate vector boson is not manifestly captured by the amplitudes provided in Ref.~\cite{Bern_1998}, but is needed for the applications described above.  In fact, up to now, 
whenever the need for such ``missing'' amplitudes arose \cite{Pellen:2021vpi,Gauld:2021ule}, they  were obtained numerically using  
capabilities of one-loop providers   \cite{Buccioni:2019sur,Actis_2017}.

A recalculation of the helicity amplitudes for the transition of a longitudinally-polarized electroweak vector boson to four partons and, especially, the simplification of the final results needed to achieve the degree of compactness as in Ref.~\cite{Bern_1998} is a daunting task. However, as we show in this paper, this is not necessary because all amplitudes for the ``missing'' polarization of the vector boson can be derived directly from the analytic amplitudes in Refs~\cite{Bern_1998,DeLaurentis:2025dxw} through a simple observation. 
Since the calculations of Refs~\cite{Bern_1998,DeLaurentis:2025dxw} were performed for arbitrary lepton momenta, subject only to the constraint that their sum equals the vector-boson momentum, the resulting amplitudes retain the full information about all polarization states of the off-shell vector boson. The procedure described in this paper shows how this information can be uncovered. 

The remainder of the paper is organized as follows.
In Section~\ref{sect:2} we make the discussion of the polarization vectors of the off-shell vector boson precise, and explain how to compute amplitudes for an arbitrary polarization of a vector boson using amplitudes calculated in Refs~\cite{Bern_1998,DeLaurentis:2025dxw} as a starting point.
In Section~\ref{sect:3} we work out the simple case of the transition of an electroweak vector boson to three jets, and demonstrate the relation between amplitudes for the transversal and the longitudinal polarizations of the vector boson.
In Section~\ref{sect:4} we discuss the $V \to 4j$ one-loop amplitudes and explain how the ``missing'' helicity amplitudes were checked in this case. 
In Section~\ref{sect:5} we present a unified analytic representation for all polarizations of the $V\to4j$ one- and two-loop leading-color finite remainders.
We conclude in Section~\ref{sect:conclusions}. All one-loop amplitudes for $V \to 3j$ and $V \to 4j$ transitions, together with the one- and two-loop leading-color finite remainders for $pp \to Vjj$, are available in computer-readable form in the ancillary files accompanying this paper.

%%%%%%%%%%%%%%%%%%%%%%%%%%%%%%%%%%%%%%%%%%%%%%%%%%%%%%%%%%%%%%
%                      SECTION 2                             %
%%%%%%%%%%%%%%%%%%%%%%%%%%%%%%%%%%%%%%%%%%%%%%%%%%%%%%%%%%%%%%

\section{Vector-boson amplitudes and polarization states}
\label{sect:2}

A prototypical loop amplitude with QCD corrections, needed to study processes such as Higgs
production in weak boson fusion in association with jets or associated Higgs
and vector boson production with additional jets, is shown in Fig.~\ref{fig1}. It can be written as
\begin{equation}\label{eq:ampitude-form-factor-decomposition}
{\cal A} =
A_{V \to 4j}^{\mu}(q)\,
D_{V}(q)\,
\rho_{\mu \nu}(q)\,
J^\nu(q) \, ,
\end{equation}
where $q$ denotes the four-momentum of an off-shell vector
boson, $A^\mu_{V \to 4j}$ describes its transition to four partons, and
$J^\mu(q)$ encodes the remaining process-dependent structures.
The propagator of the intermediate vector boson is
\begin{equation}
D_V(q) = \frac{1}{q^2-M_V^2
+{\rm i} \, \Gamma_V M_V} \, ,
\end{equation}
and the polarization density matrix of the off-shell vector boson reads
\begin{equation}
\rho_{\mu \nu}(q) =
- g_{\mu \nu}
+ \frac{q_\mu q_\nu}{q^2} \, .
\label{eq2.3}
\end{equation}

A common approach to describe five-point one-mass amplitudes, which allows to exploit massless spinor-helicity techniques, is to express them in terms of six-point massless amplitudes. This corresponds to selecting a specific current $J^\mu(q)$ representing the transition $V\rightarrow \bar\ell \ell$ (which crucially may not be the one of interest). In this notation, two \emph{arbitrary} light-like four-momenta $p_5$ and $p_6$ that add up to $q$ are introduced
\begin{equation}
p_{5}^2=p_{6}^2 = 0 \, , \; q = p_5 + p_6 \, .
\end{equation}
In terms of $p_5$ and $p_6$, we can explicitly relate the polarization density matrix $\rho_{\mu\nu}$ to three polarization vectors that, for a specific choice of 
$p_{5,6}$, will correspond to the \emph{physical polarization vectors of the massive vector boson}. 

Specifically, it is easy to check that the density matrix in Eq.~(\ref{eq2.3}) satisfies the following equation 
\be
    \rho^{\mu \nu} = \sum_{\lambda = \pm, L} \, \varepsilon_q^{(\lambda),\mu} \, \varepsilon_q^{(-\lambda),\nu} \, ,
    \label{eq2.5}
\ee
where the polarization vectors can be chosen  
as we describe below. 
The two ``transverse'' polarizations read
\begin{equation}
    \varepsilon^{(+),\mu}_{q} = \frac{\langle6|\gamma^{\mu}|5]}{\sqrt{2} \, \langle 6 5 \rangle} \, , \quad 
    \varepsilon^{(-),\mu}_{q} = -\frac{[6|\gamma^{\mu}|5\rangle}{\sqrt{2} \, [6 5]} \, . 
    \label{eq2.6}
\end{equation}
They match the usual positive and negative helicity polarization states for a massless 
particle with momentum $p_5$ 
and reference momentum $p_6$. 
The ``longitudinal'' polarization vector can be deduced from Eq.~(\ref{eq2.5}) using the polarization vectors $\varepsilon^{(\pm)}_q$. One finds 
\begin{equation}
    \varepsilon_{q}^{(L),\mu} = \frac{p_5^\mu - p_6^\mu}{\langle 5 6 \rangle } \, , \quad
    \varepsilon_{q}^{(-L),\mu} = \frac{p_6^\mu - p_5^\mu}{[5 6]} \, ,
\end{equation}
or, equivalently, in spinor-helicity notation
\begin{equation}\label{eq:massless-spinor-longitudinal-polarization}
    \varepsilon_{q}^{(L),\mu}
    = \frac{[5|\gamma^\mu|5\rangle - [6|\gamma^\mu|6\rangle}{2 \, \langle 5 6 \rangle} \, , \quad 
    \varepsilon_{q}^{(-L),\mu}
    = \frac{[6|\gamma^\mu|6\rangle - [5|\gamma^\mu|5\rangle}{2 \, [5 6]}  \, .
\end{equation}

%%%%%%%%%%%%%%%%%%%%%%%%%%%%%%%%%%%%%%%
% FIGURE
\FigOneLoopJdiag{t}{\label{fig1}
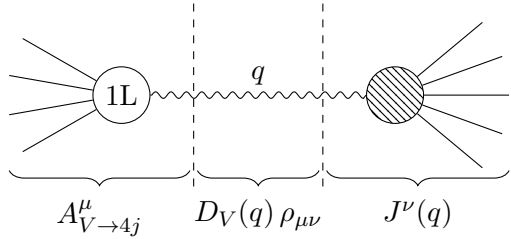
The embedment of a  one-loop amplitude $A^\mu_{V\rightarrow 4j}$  within an  amplitude 
of a more general process. See the text for 
details. 
}
%%%%%%%%%%%%%%%%%%%%%%%%%%%%%%%%%%%%%%%

Gauge invariance implies that the amplitude satisfies the Ward identity
\begin{equation}
{\cal A}_{V \to 4j}^\mu(q) \, q_\mu = 0 \, ,
\end{equation}
while the polarization vectors are constructed to be ortho-normal and satisfy the following  relations 
\begin{align}
\varepsilon^{(\lambda),\mu} \, \varepsilon^{(-\lambda')}_\mu = -\delta^{(\lambda, \lambda')} \; , \qquad \varepsilon^{(\lambda),\mu} \, q_\mu = 0 \, .
\end{align}

Amplitudes shown  in Eq.~\eqref{eq:ampitude-form-factor-decomposition} with the current $J^\mu(q)$ replaced with polarization vectors 
$\varepsilon^{(\pm)}_q $ in Eq.~(\ref{eq2.6})  
were calculated at one loop in Ref.~\cite{Bern_1998}, while the two-loop amplitudes in the leading-color approximation were first obtained in Ref.~\cite{Abreu:2021asb} and then simplified in Ref.~\cite{DeLaurentis:2025dxw}. 
We note that in  Ref.~\cite{Abreu:2021asb} the helicity amplitude $\mathcal{A}$ was obtained by first calculating three form factors corresponding to the three physical degrees of freedom of the massive vector boson  and then  contracted with the polarization vectors 
$\varepsilon^{(\pm)}_q $.

The goal of this paper is to point out that, when  interpreted appropriately, the six-point computations retain the full information about the three physical polarization states of the massive vector boson or, equivalently, that a single form factor for any of the three polarization states for the massive vector bosons suffices to fully determine the amplitude. This is best understood in terms of the spin-covariance of the amplitude.

%%%%%%%%%%%%%%%%%%%%%%%%%%%%%%%%%%%%%%%%%%%%%%%%%%%%%%%%%%%%%%
%                      SCETION 2.1                           %
%%%%%%%%%%%%%%%%%%%%%%%%%%%%%%%%%%%%%%%%%%%%%%%%%%%%%%%%%%%%%%

\subsection{Little-group covariance in the massive spinor-helicity formalism}

The amplitude involving an off-shell vector boson can be conveniently
described using the massive spinor-helicity (spin-spinor) formalism
\cite{Conde:2016izb, Conde:2016vxs, Arkani-Hamed:2017jhn, Ochirov:2018uyq, Shadmi:2018xan}.
In this approach the momentum of the massive particle is written as
\be\label{eq:massless-to-massive-spinors}
|\boldsymbol q^I \rangle [\boldsymbol q_I|
=
|\boldsymbol q^1 \rangle [\boldsymbol q_1|
+
|\boldsymbol q^2 \rangle [\boldsymbol q_2|
=
|5\rangle[5| + |6\rangle[6| .
\ee
The above equation elucidates the relation between the five-point one-mass
(``bold'') notation and the six-point massless notation.

The index $I=1,2$ labels the $\mathrm{SU}(2)$ little group of a massive
particle. In the massive spinor-helicity formalism the spinors transform as
\begin{equation}
|\boldsymbol q^I\rangle \;\rightarrow\;
U^I{}_J\,|\boldsymbol q^J\rangle,
\qquad
[\boldsymbol q_I| \;\rightarrow\;
[\boldsymbol q_J|\,(U^{-1})^J{}_I,
\qquad U\in \mathrm{SU}(2),
\end{equation}
while the momentum
$p_{\alpha\dot\alpha}=|\boldsymbol q^I\rangle[\boldsymbol q_I|$
remains invariant.
Scattering amplitudes therefore transform covariantly under the
massive little group and naturally carry $\mathrm{SU}(2)$ (spin)
indices. For a spin-1 particle the amplitude transforms as
\begin{equation}
A^{IJ} \;\rightarrow\;
U^I{}_K\,U^J{}_L\,A^{KL} \, .
\end{equation}

In the bold notation, the polarization tensor takes the form
\be\label{eq:spin-spinor-polarization-tensor}
\varepsilon^{\mu,IJ}_q
   \;=\;
   \frac{1}{\sqrt{2}\,m}\,
   [ \boldsymbol q^I \,|\, \gamma^\mu \,|\, \boldsymbol q^J \rangle \, ,
\ee
with only three independent combinations corresponding to physical
polarizations. These combinations are naturally identified through the
spin-one Clebsch--Gordan decomposition,
\begin{align}
\bigl|1,+1\bigr\rangle &= \bigl|\tfrac12,+\tfrac12\bigr\rangle
                         \otimes \bigl|\tfrac12,+\tfrac12\bigr\rangle \, , \\[6pt]
\bigl|1,0\bigr\rangle &= \frac{1}{\sqrt{2}}
  \Bigl(
      \bigl|\tfrac12,+\tfrac12\bigr\rangle\!
      \otimes\!
      \bigl|\tfrac12,-\tfrac12\bigr\rangle
      \;+\;
      \bigl|\tfrac12,-\tfrac12\bigr\rangle\!
      \otimes\!
      \bigl|\tfrac12,+\tfrac12\bigr\rangle
  \Bigr) \, , \\[6pt]
\bigl|1,-1\bigr\rangle &= \bigl|\tfrac12,-\tfrac12\bigr\rangle
                         \otimes \bigl|\tfrac12,-\tfrac12\bigr\rangle \, .
\end{align}
In terms of Eq.~\eqref{eq:spin-spinor-polarization-tensor},
the corresponding polarization vectors are given by \cite{Wu:2021nmq}
\begin{align}
\varepsilon^{(+),\mu}_q \; &=\; \varepsilon^{\mu, 22} \, , \\
\varepsilon^{(L), \mu}_q \; &=\; \frac{1}{\sqrt{2}}\!\left(\varepsilon^{\mu, 12} + \varepsilon^{\mu, 21}\right) \, , \label{eq:spin-spinor-longitudinal-polarization} \\
\varepsilon^{(-), \mu}_q \; &=\; \varepsilon^{\mu, 11} \, .
\end{align}
This shows that the $A^{(+)},A^{(-)}$ and $A^{(L)}$ amplitudes are not independent but rather correspond to  particular choices of spin indices $I, J$ for the spin-covariant amplitude $A^{IJ}$. Amplitudes contracted with the  decay current $[5 | \gamma^\mu | 6 \rangle $ can then be identified with the $(\varepsilon^\mu)^{\;2}_{1}$ component.

The equivalence of the definitions in Eqs~\eqref{eq:spin-spinor-longitudinal-polarization} and \eqref{eq:massless-spinor-longitudinal-polarization} can be shown explicitly. To this end, we first expand the definition in Eq.~\eqref{eq:spin-spinor-longitudinal-polarization},
\be
\varepsilon^{(L), \mu}_q
   \;=\;
   \frac{1}{2\,m}\,\left(
   [ \boldsymbol q^1 \,|\, \gamma^\mu \,|\, \boldsymbol q^2 \rangle + 
   [ \boldsymbol q^2 \,|\, \gamma^\mu \,|\, \boldsymbol q^1 \rangle\right).
   \label{eq2.21}
\ee
Then,  lowering the index with the help of the $\mathrm{SU}(2)$ spin metric, i.e.~the Levi-Civita tensor,
\be
[\boldsymbol q^I| = \epsilon^{IJ} [\boldsymbol q_J| \; \Rightarrow \;  
[\boldsymbol q^1| = \epsilon^{12} [\boldsymbol q_2| = [\boldsymbol q_2| \;\, \text{and} \;\, [\boldsymbol q^2| = \epsilon^{21} [\boldsymbol q_1| = -[\boldsymbol q_1|, 
\ee
we arrive at Eq.~\eqref{eq:massless-spinor-longitudinal-polarization}.

The mass $m$ in Eq.~(\ref{eq2.21})
%dependence of the denominator 
can be analytically continued to complex kinematics by using the holomorphic and anti-holomorphic masses 
\be
\langle \boldsymbol q^I | \boldsymbol q^J \rangle = \mathfrak{m} \epsilon^{JI} ,
\ee
\be
[ \boldsymbol q^I | \boldsymbol q^J ] =  \mathfrak{\bar m} \epsilon^{IJ} ,
\ee
which gives $\mathfrak{m} \, \mathfrak{\bar m} = m^2$ but avoids introducing square roots.
This feature may be important in various circumstances, for example when working over finite fields. In this notation, $\varepsilon^{(+)}_q$ and $\varepsilon^{(L)}_q$ have a factor of $\mathfrak{m}$, and $\varepsilon^{(-)}_q$ and $\varepsilon^{(-L)}_q$ a factor of $ \mathfrak{\bar m}$.

\subsection{A trick to recover the longitudinal polarization}

Following the discussion in  the previous section, here we show how to extract the amplitudes for the ``longitudinally-polarized'' vector boson from amplitudes  for a ``transversely-polarized'' vector boson, 
e.g. the ones computed in Ref.~\cite{Bern_1998} for $e^+e^- \to 4j$ at one loop. For definiteness, we consider the polarization vector $\varepsilon_{q}^{(+)}$. Then, available amplitudes read 
\begin{equation}
    A_{V \to 4j}^{\mu} \, \varepsilon^{(+)}_{q,\mu}
    = \frac{[5|{\cal O}|6\rangle}{\sqrt{2} \, [5 6]} \, , 
\label{AmpEpsPlus}
\end{equation}
where ${\cal O}$ indicates an operator composed of various spinors, etc. Each independent amplitude is characterized by its own operator ${\cal O}$, but they all have one feature in common, namely, they do not depend on the momenta $p_{5,6}$ or the corresponding spinors.
The only dependence of ${\cal O}$ on $p_{5,6}$ that is admissible is the dependence on $q = p_5 + p_6$.

Although the amplitudes presented in Refs~\cite{Bern_1998,DeLaurentis:2025dxw} appear to have a more complex dependences on $|5]$ and $|6\rangle$ than what is shown in Eq.~(\ref{AmpEpsPlus}), one can rewrite them as in that equation. 
This follows from the fact that in the calculation of Refs~\cite{Bern_1998,DeLaurentis:2025dxw} momenta $p_{5,6}$ were kept arbitrary, except for being light-like vectors that satisfy the momentum conservation relation $q = p_5 + p_6$. 
Thus, except for the constraint $\varepsilon^{(+), \mu}_{q} \, q_{\mu} = 0$, one cannot introduce hidden dependences on $p_{5,6}$ in the amplitudes computed in Refs~\cite{Bern_1998,DeLaurentis:2025dxw} through simplifications. This implies that one can always rewrite  amplitudes presented in Refs~\cite{Bern_1998,DeLaurentis:2025dxw} in the form shown in Eq.~(\ref{AmpEpsPlus}).
Once this is done, we can read off the operator ${\cal O}$ from an expression for a particular helicity amplitude.

Then, given the operator ${\cal O}$ for an amplitude as in Eq.~(\ref{AmpEpsPlus}) and taking into account the proximity of the polarization vectors for $\pm$ and $L$ polarizations as well as  the fact that $p_{5,6}$ were kept generic in 
computations reported in Refs~\cite{Bern_1998,DeLaurentis:2025dxw}, we conclude that the commensurate  amplitude for    the ``longitudinally-polarized'' vector boson reads 
\begin{equation}\label{eq:longitudinal-amplitude}
    A_{V \to 4j}^\mu \, \varepsilon^{(L)}_{q,\mu}
    = \frac{[5|{\cal O}|5\rangle - [6|{\cal O}|6\rangle}{2 \, \langle 5 6 \rangle} \, .
\end{equation}
Obviously, it is crucial  that the operator ${\cal O}$ in the above equation and in Eq.~(\ref{AmpEpsPlus}) is determined for fixed helicities of all QCD partons, and that $\varepsilon^{(L), \mu}_{q} \, q_{\mu} = 0$
is satisfied, to respect simplifications introduced by the constraint $\varepsilon^{(+), \mu}_{q} \, q_{\mu} = 0$ in the computation of amplitudes 
for the transverse polarization. 

We note that the very  fact that 
terms proportional to $q^\mu$ in 
${\cal A}_{V \to 4j}^\mu$ are projected away, when it is contracted with the physical polarizations, leads to difficulties when trying to verify the Ward identity
for the operator ${\cal O}$,
\begin{equation}\label{eq:ward-identity-six-point}
    [5|{\cal O}|5\rangle + [6|{\cal O}|6\rangle = 0 \, .
\end{equation}
We have found that  this equation holds true in the simplest cases, for example for   tree amplitudes, but not  in more complicated ones. The reason is that the operator  $\mathcal{O}$ can be reconstructed only up to  terms that vanish when contracted with physical polarizations, and these terms are essential for 
the fulfillment of the Ward identity. 

To illustrate this further, we consider a simple example where the massless six-point expression reads
\be
\frac{⟨i6⟩⟨j6⟩}{⟨56⟩} = \frac{⟨i6⟩⟨j6⟩[65]}{⟨56⟩[65]}.
\ee
Removing the propagator and focussing on the numerator we find
\be
⟨i6⟩⟨j6⟩[65] = [5|5+6|i⟩⟨j6⟩ = [5|5+6|j⟩⟨i6⟩.
\ee
It follows that $\mathcal{O}$ can be taken to be either $|5+6|i\rangle\langle j|$
or $|5+6|j\rangle\langle i|$, or more generally any linear combination
\begin{equation}
\mathcal{O}
=
a\,|5+6|i\rangle\langle j|
+
(1-a)\,|5+6|j\rangle\langle i|\, ,
\end{equation}
Crucially, any such choice leads to the same result in the asymmetric combination of Eq.~\eqref{eq:longitudinal-amplitude}.
The ambiguity corresponds precisely to terms proportional to
$q=p_5+p_6$\footnote{Trees work out automatically because $i=j$.}
\begin{equation}
|5+6|i\rangle\langle j|
-
|5+6|j\rangle\langle i|
=
q_{\dot\alpha\beta}
\big(
\lambda_i^{\beta}\lambda_{j\alpha}
-
\lambda_j^{\beta}\lambda_{i\alpha}
\big)
=
\langle j\,i\rangle\,q_{\dot\alpha\alpha}.
\end{equation}
However, if we require $\mathcal{O}$ to satisfy the Ward identity
\eqref{eq:ward-identity-six-point}, we must choose the symmetric
combination
\begin{equation}
\mathcal{O}
=
\frac12
\big(
|5+6|i\rangle\langle j|
+
|5+6|j\rangle\langle i|
\big),
\end{equation}
since
\begin{equation}
q^{\alpha\dot\alpha}\,
q_{\dot\alpha\beta}
\big(
\lambda_i^{\beta}\lambda_{j\alpha}
+
\lambda_j^{\beta}\lambda_{i\alpha}
\big)
=
q^2
\big(
\langle i\,j\rangle + \langle j\,i\rangle
\big)
=
0 .
\end{equation}

In terms of the spin-spinor formalism, the ambiguity in $\mathcal{O}$ translates into an ambiguity in spin-covariant amplitude
\be\label{eq:spin-covariant-amplitude}
\mathcal{A}^\mu \, \epsilon_\mu^{IJ} = A^{IJ} = r_i^{IJ} \, I_i \, ,
\ee
where $r_i^{IJ}$ are rational functions carrying spin indices and $I_i$ are Feynman integrals. Each $ r_i^{IJ}$ is recovered only up to a transpose. Irrespectively of the choice, we obtain the correct longitudinal amplitude $A^L = (A^{12} + A^{21})/\sqrt{2}$, but only after symmetrization of each $r_i^{IJ}$ we reproduce $0 = (A^{12} - A^{21})$.

%%%%%%%%%%%%%%%%%%%%%%%%%%%%%%%%%%%%%%%%%%%%%%%%%%%%%%%%%%%%%%
%                        SECTION 3                           %
%%%%%%%%%%%%%%%%%%%%%%%%%%%%%%%%%%%%%%%%%%%%%%%%%%%%%%%%%%%%%%

\section{One-loop amplitudes for  \texorpdfstring{$V \to 3j$}{}}
\label{sect:3}

We will illustrate the procedure described in the previous section by considering the case of a vector boson decaying to \emph{three} QCD partons. The relevant amplitudes are presented in Appendix IV of Ref.~\cite{Bern_1998}. They are computed in dimensional regularization where the dimension of space-time $d$ is parametrized in the standard way,  $d = 4 -2 \ep$. 

For definiteness, we consider the leading-color amplitude $A_{5}(1_{q}^{+},2^{+},3_{\bar{q}}^{-}; 5_{\bar{e}}^-,6_{e}^{+})$; see Eqs (IV.1--IV.5) of Ref.~\cite{Bern_1998}. It reads 
\begin{equation}
    A_{5}(1_{q}^{+},2^{+},3_{\bar{q}}^{-},5_{\bar{e}}^-,6_{e}^{+})
    = c_\Gamma \left[ A^{5}_{\rm tree} \, V_{\rm div} + \rm{i} \, F_{\rm fin} \right] \,,
\end{equation}
where
\begin{equation}
        c_\Gamma = \frac{\Gamma(1+\ep) \, \Gamma^{2}(1-\ep)}{(4 \pi)^{2-\ep} \, \Gamma(1-2\ep)} \,, \\
\end{equation}
and 
\begin{align} 
    A^{5}_{\rm tree}
    = & -\rm{i} \, \frac{\langle 3 5 \rangle^{2}}{\langle 1 2 \rangle \langle 2 3 \rangle \langle 5 6\rangle} \, , \\
    V_{\rm div}
    = & - \frac{1}{\ep^2} \left[\left( \frac{\mu^2}{-s_{12}} \right)^{\!\!\ep}
    + \left( \frac{\mu^2}{-s_{23}} \right)^{\!\!\ep} \right]
    - \frac{3}{2\ep} \left( \frac{\mu^2}{-s_{23}} \right)^{\!\!\ep} - 3 \, , \\
    F_{\rm fin}
    = &~ \frac{\langle 3 5 \rangle ^2}{\langle 1 2 \rangle \langle 2 3 \rangle \langle 5 6\rangle} \,
    {\rm Ls}_{-1} \left( \frac{-s_{12}}{-s_{56}} \, , \ \frac{-s_{23}}{-s_{56}} \right) 
    -\frac{\langle 3 5 \rangle \langle 3|16|5 \rangle}{\langle 1 2 \rangle \langle 2 3 \rangle \langle 5 6 \rangle} \,
    \frac{{\rm L}_{0}\left( \frac{-s_{23}}{-s_{56}} \right)}{s_{56}}  \\
    & + \frac{1}{2} \frac{\langle 3|16|5 \rangle^2}{\langle 1 2 \rangle \langle 2 3 \rangle \langle 5 6 \rangle} \,
    \frac{{\rm L}_{1}\left( \frac{-s_{23}}{-s_{56}} \right)}{s_{56}^2} \, . \notag
\end{align}
Functions ${\rm Ls}_{-1}, {\rm L}_0$ and ${\rm L}_1$ are combinations of logarithms and polylogarithms; they are defined in Appendix II of Ref.~\cite{Bern_1998}.

Our goal is to turn this amplitude into an amplitude for the longitudinally-polarized vector boson, as defined in Section~\ref{sect:2}.
The first step is to write the amplitude $A_5$ in the form $A_5 \sim [6|{\cal O} |5 \rangle$ where the operator ${\cal O}$ does not depend on $p_5$ and $p_6$ separately, but only on their sum. Since functions ${\rm Ls}_{-1}$, etc. depend on $s_{56} = (p_5 + p_6)^2$, we only need to manipulate the various spinor chains in the amplitude $A_5$. 

We will start with the tree amplitude $A_{\rm tree}^5$. To bring it to the desired form, we multiply it with $1 = [65]/[65]$, and use $\langle 5 6 \rangle [6 5] = s_{56}$ in the denominator. We find   
\begin{equation}
    -\rm{i} \, \frac{\langle 3 5 \rangle^2}{\langle 1 2 \rangle \langle 2 3 \rangle \langle 56 \rangle}
    = \rm{i} \, \frac{[65] \langle 5 3 \rangle \langle 3 5 \rangle}{\langle 1 2 \rangle \langle 2 3 \rangle \, s_{56}}
    = - \rm{i} \, \frac{[6|1+2|3 \rangle \langle 3 5 \rangle}{\langle 1 2 \rangle \langle 2 3 \rangle \, s_{56}} \, ,
\end{equation}
where in the last step we have used momentum conservation and the Dirac equation to write 
\begin{equation}
    [65] \langle 5 3 \rangle
    = [6|5|3\rangle = [6|6+5+3|3\rangle
    = -[6|1+2|3\rangle \, .
\end{equation}
Thus,  in this case we identify the operator ${\cal O}$  with 
\begin{equation}
    - \rm{i} \, \frac{| 1 + 2 |3 \rangle \langle 3|}{\langle 1 2 \rangle \langle 2 3 \rangle \, s_{56}} \, , 
\end{equation}
and it is now straightforward to take matrix elements of this operator with respect to $[5| \, , |5 \rangle$ and $[6| \, , |6 \rangle$ spinors, constructing the term contributing to the amplitude of the longitudinally-polarized vector boson decay 
to three QCD partons. 

Spinors in the second term in $F_{\rm fin}$ are re-written as follows 
\begin{equation}
    \frac{\langle 3 5 \rangle \langle 3|16| 5 \rangle}{\langle 1 2 \rangle \langle 2 3 \rangle \langle 5 6\rangle}
    = -\frac{[6|1|3\rangle \langle 3 5\rangle }{\langle 1 2 \rangle \langle 2 3 \rangle} \, .
\end{equation}
This is the matrix element of the operator 
\begin{equation}
  - \frac{| 1 |3 \rangle \langle 3|}{\langle 1 2 \rangle \langle 2 3 \rangle} \, ,
\end{equation}
evaluated between spinors $[6|$ and $|5 \rangle$.
 
Finally, the spinor part of the last term in the expression for $F_{\rm fin}$ is written as 
\begin{equation}
    \frac{\langle 3|16|5 \rangle^2}{\langle 1 2 \rangle \langle 2 3 \rangle \langle 5 6 \rangle}
    = - \frac{[6|1|3 \rangle \langle 3|16|5 \rangle}{\langle 1 2 \rangle \langle 2 3 \rangle}
    = - \frac{[6|1|3 \rangle \langle 3|1|5+6|5 \rangle}{\langle 1 2 \rangle \langle 2 3 \rangle} \, .
\end{equation}
We recognize the matrix element of the following operator 
\begin{equation}
    - \frac{| 1 |3 \rangle \langle 3| 1 | 5 + 6|}{\langle 1 2 \rangle \langle 2 3 \rangle} \, ,
\end{equation}
computed between spinors $[6|$ and $|5 \rangle$.

Combining the different contributions, we obtain 
\begin{equation}
    A^{5}(1_{q}^{+},2^{+},3_{\bar{q}}^{-},5_{\bar{e}}^-,6_{e}^{+})
    = c_\Gamma \, [6|{\cal O}(1_{q}^{+},2^{+},3_{\bar{q}}^{-})|5 \rangle \, ,
\label{V3jHelAmpL}
\end{equation}
where 
\begin{equation}
\begin{gathered}
    {\cal O}(1_{q}^{+},2^{+},3_{\bar{q}}^{-})
    = - \rm{i} \, \frac{| 1 +  2 |3 \rangle \langle 3|}{\langle 1 2 \rangle \langle 2 3 \rangle \, s_{56}} \,
    \left[ V_{\rm div} - {\rm Ls}_{-1} \left( \frac{-s_{12}}{-s_{56}} \, , \, \frac{-s_{23}}{-s_{56}} \right) \right] \\
    + \rm{i} \, \frac{| 1 |3 \rangle \langle 3|}{\langle 1 2 \rangle \langle 2 3 \rangle} \,
    \frac{{\rm L}_{0}\left( \frac{-s_{23}}{-s_{56} } \right)}{s_{56}}
    - \rm{i} \, \frac{1}{2} \frac{| 1 |3 \rangle \langle 3| 1 | 5 + 6|}{\langle 1 2 \rangle \langle 2 3 \rangle} \,
    \frac{{\rm L}_{1} \left( \frac{-s_{23}}{-s_{56}} \right)}{s_{56}^2} \, .
\end{gathered} 
\end{equation}

The amplitude  for the longitudinally-polarized vector boson is then obtained by computing $[5|{\cal O}|5 \rangle$ and $[6|{\cal O}|6 \rangle$, and taking their difference, as explained in the previous section. We note that in the calculation of Ref.~\cite{Bern_1998}, the propagator $ 1 / s_{56}$ is included in the definition of the amplitudes.

One- and two-loop helicity amplitudes for $V \to 3j$ transition for an arbitrary polarization of the vector boson have been known for quite some time; see Refs~\cite{Garland_2002,Gehrmann_2023}. We used these results to check amplitudes for the longitudinal polarization of the vector boson obtained following the procedure described above. 

We note that the amplitudes in Ref.~\cite{Bern_1998} are computed in the four-dimensional helicity (FDH) scheme \cite{Bern_1991aq}, whereas the results of Refs~\cite{Garland_2002,Gehrmann_2023} are quoted in the 't Hooft-Veltman (tHV) scheme \cite{tHV1973}. This difference in the regularization schemes can be accounted for by employing universal shifts proportional to tree amplitudes \cite{Kunszt_1994}.
We provide all the spin-covariant primitive amplitudes for $V \to 3j$, computed in the FDH scheme and without the propagator $1/s_{56}$, in the ancillary file \texttt{Results\textunderscore V3j.m}.
%%%%%%%%%%%%%%%%%%%%%%%%%%%%%%%%%%%%%%%%%%%%%%%%%%%%%%%%%%%%%%
%                        SECTION 4                           %
%%%%%%%%%%%%%%%%%%%%%%%%%%%%%%%%%%%%%%%%%%%%%%%%%%%%%%%%%%%%%%

\section{\texorpdfstring{One-loop amplitudes for $V \to 4j$}{}}
\label{sect:4}

To obtain helicity amplitudes for the $V \to 4j$ transition for the longitudinally-polarized vector boson, we follow the procedure described in the previous sections. The calculation is straightforward, albeit tedious. However, checking the obtained results is difficult because no one-loop $V \to 4 j$ amplitudes for arbitrary vector-boson polarization 
%beyond Ref.~\cite{Bern_1998} 
seem to  be publicly available. Hence, for the purpose of the checks, we had to compute these amplitudes independently. We describe the main steps involved in their calculation in this section. 

For definiteness, we consider two processes 
\begin{equation}
   0 \to  V(p_{V}) \, q(p_{1}) \, \bar{q}(p_{2}) \, g(p_{3}) \, g(p_{4}) \, ,
\label{Vqqbgg}
\end{equation}
and
\begin{equation}
    0 \to  V(p_{V}) \, q(p_{1}) \, \bar{q}(p_{2}) \, Q(p_{3}) \, \bar{Q}(p_{4}) \, ,
\label{VqqbQQb}
\end{equation}
and assume that these transitions are  facilitated by the vector current $\bar q \gamma_\mu q$. All final-state partons are massless, so that their momenta are light-like $p_{i}^{2} = 0$ for $i=1,2,3,4$ and momentum conservation implies $p_{V}+p_{1}+p_{2}+p_{3}+p_{4}=0$. 

In Ref.~\cite{Bern_1998} color-ordered tree and one-loop primitive amplitudes for the processes in \crefrange{Vqqbgg}{VqqbQQb} are presented. If one starts from conventional Feynman diagrams and uses conventional Feynman rules, it is possible to identify contributions of Feynman diagrams to some of the primitive amplitudes through their dependences on the generators of the $\textrm{SU}(3)$ gauge group, while in other cases  the correspondence between Feynman diagrams and linear combinations of primitive amplitudes can be established. Be it as it may, we deal with color degrees of freedom in the standard way and remove them in the early steps of the calculation. 
 
Once the color degrees of freedom are removed, amplitudes are written as linear combinations of independent Lorentz structures $T_{i}$ with coefficients $F_{i}$ that depend on the scalar products between the four-momenta of all partons involved in the process
\begin{equation}
    {\cal{A}} \sim \sum_{i=1}^{N} F_{i} \, T_{i} \, .
\end{equation}

To build independent Lorentz-invariant structures, we invoke the discussion in Ref.~\cite{Peraro_2021} where 
simplifications owing to the four-dimensional nature of the momenta and polarization vectors of external partons are described. In short, using linearly-independent four-momenta $p^\mu_{1,2,3,4}$ as basis vectors and the fact that the helicity along any fermion line is conserved in massless QCD, it is easy to see that for the process $V \to q\bar{q}gg$ of Eq.~(\ref{Vqqbgg}), there are 24 independent tensors structures.
We choose them as follows  
\begin{equation}
    T \sim  \Bar{u}(p_1)\, \slashed{p}_{3,4} \, u(p_{2}) \, (\varepsilon_{3} \cdot p_{1,2}) \, (\varepsilon_{4} \cdot p_{1,2}) \, (\varepsilon_{V} \cdot p_{1,2,3}) \, .
\label{Tqqbgg}
\end{equation}

Here, $\varepsilon_{i=3,4}$ and $\varepsilon_{V}$ are the polarization vectors of the gluons and the virtual vector boson, respectively. We note that we assumed that all polarizations are physical 
\begin{equation}
    \varepsilon_3 \cdot p_3 = \varepsilon_4 \cdot p_4 = 0 \, , \quad \varepsilon_V \cdot p_V = 0 \, , 
\end{equation}
and employed the gauge-fixing conditions $\varepsilon_{3} \cdot p_{4} = \varepsilon_{4} \cdot p_{3} = 0$ to reduce the number of independent Lorentz structures. 

For the process $V \to q\bar{q}Q\bar{Q}$ of Eq.~(\ref{VqqbQQb}) there are 12 independent tensor structures, which we choose to be 
\begin{equation}
    T \sim  \Bar{u}(p_1)\, \slashed{p}_{3,4} \, u(p_{2}) \, \Bar{u}(p_3)\, \slashed{p}_{1,2} \, u(p_{4}) \, (\varepsilon_{V} \cdot p_{1,2,3}) \, .
\label{TqqbQQb}
\end{equation}

To calculate the form factors, we take the hermitian conjugate of each element of the tensor basis $T_i \to T_i^{\dagger}$, and write 
\begin{equation}
    \sum_{\rm pols} T_{i}^{\dagger} \, {\cal A}
    = \sum_{j} M_{ij} \, F_{j} \, ,
\end{equation}
where on the left-hand side the summation over polarizations of all partons and the vector boson is performed, and on the right-hand side the matrix $M_{ij}$ is defined as follows
\begin{equation}
    M_{ij} = \sum_{\text{pols}} \, T_{i}^{\dagger} \, T_{j} \, .
\end{equation}

Using  the inverse matrix $c_{ij} = M^{-1}_{ij}$, we find 
\begin{equation}
    F_{i} = \sum_{j} \, c_{ij} \, \sum_{\rm pols} \, T_{j}^{\dagger} \, {\cal A} \, .
\end{equation}
Therefore, the operator acting on the amplitude ${\cal A}$ on the right-hand side of the above equation is the projection operator on the form factor $F_{i}$ associated with the Lorentz structure $T_{i}$. We provide two  matrices $c_{ij}$ for the processes  in Eqs~(\ref{Vqqbgg}--\ref{VqqbQQb}) in the ancillary files texttt{cijVqqbgg.m} and \texttt{cijVqqbQQb.m}. We note that they were computed with \texttt{FiniteFlow}~\cite{Peraro_2019}.

Once projection operators are constructed, the calculation of the form factors proceeds in the standard way. We use \texttt{FeynArts}~\cite{Hahn_2001} to generate Feynman diagrams, and \texttt{FeynCalc}~\cite{Mertig_1991,Shtabovenko_2016,Shtabovenko_2020,Shtabovenko_2025} to remove color degrees of freedom. Diagrams are further processed using \texttt{FORM}~\cite{Kuipers_2013}, where the projection operators are applied.
Subsequently, the output is written in terms of scalar integrals in \texttt{Mathematica}, which are then reduced to master integrals using \texttt{Kira}~\cite{Maierh_fer_2018,Klappert_2021,lange2025}. 
Finally, the one-mass pentagon integrals are calculated using dimensional recursion relations \cite{Bern_1993,Bern_1994}, and every master integral is checked numerically  with \texttt{AMFlow}~\cite{Liu_2023}.

It is useful to note that the approach described above leads to huge algebraic expressions at intermediate stages of the calculation. 
Since the primary goal of this brute-force computation is to check the approach described in Section~\ref{sect:2}, we do not need to put too much effort into the simplification of the results. For this reason, we decided to perform only basic simplifications, and also used semi-numerical computations  for a collection of phase-space points.
We stress  that for numerical checks, \emph{rational} values of external momenta are chosen, so that the result is written as a linear combination of master integrals with \emph{exact} rational coefficients. 

We can check this procedure by comparing its results with the helicity amplitudes reported in Ref.~\cite{Bern_1998} from which we removed the photon propagator.\footnote{All relevant amplitudes can be found in a computer-readable form at \url{https://s3df.slac.stanford.edu/people/lance/Zqqgg_paper.html}.} In fact, our calculation allows us to compare the coefficients of all independent polylogarithmic and logarithmic functions, used in Ref.~\cite{Bern_1998} to present the amplitudes, with the results of our computation for a few phase-space points. We stress that the form factors were computed in the tHV scheme, while the FDH scheme was used in the calculation of Ref.~\cite{Bern_1998}. We account for this difference by using universal shifts proportional to tree amplitudes; see Eqs (6.4--6.5) of Ref.~\cite{Bern_1998}. Since the form factors are independent of the external polarizations, the above comparison provides a very strong check on the validity of the calculation based on the projection operators. 

Having verified the form factors, we proceed with the comparison of the helicity amplitudes for the $V \to 4j$ transition for the longitudinally-polarized vector boson. 
We construct these amplitudes from the helicity amplitudes in Ref.~\cite{Bern_1998} using the method discussed in the previous sections. We then use the set up based on the projection operators to thoroughly check them.
We emphasize  one more time that the comparison is done for the \emph{rational} external momenta, so that \emph{exact} agreement between the two results can be claimed. 
We provide all the primitive amplitudes in the FDH scheme for the $V \to 4j$ transition in the spin-covariant notation in the ancillary file \texttt{Results\textunderscore V4j.m}. Additional information on how to use them can be found in the ancillary file \texttt{README.txt}.  Numerical reference values for various helicity amplitudes are provided  
in Appendix~\ref{app:a}.
%%%%%%%%%%%%%%%%%%%%%%%%%%%%%%%%%%%%%%%%%%%%%%%%%%%%%%%%%%%%%%
%                        SECTION 5                           %
%%%%%%%%%%%%%%%%%%%%%%%%%%%%%%%%%%%%%%%%%%%%%%%%%%%%%%%%%%%%%%

\section{\texorpdfstring{Two-loop remainders for $V \to 4j$}{}}
\label{sect:5}

In this section, we demonstrate that the physical understanding and the
computational method presented in the previous sections are independent of the loop order. To this end, we explicitly show that the results of
Ref.~\cite{DeLaurentis:2025dxw}, calculated for \emph{one} out of three polarization components, contain the same amount of information as the off-shell form factor computation of Ref.~\cite{Abreu:2021asb}. 

We proceed as described in Sections~\ref{sect:3} and \ref{sect:4} for the one- and two-loop helicity remainders of the processes
\be
0 \rightarrow 
\bar q^{h_1}(p_1)\,
g^{h_2}(p_2)\,
g^{h_3}(p_3)\,
q^{h_4}(p_4)\,
\bar \ell^+(p_5)\,
\ell^-(p_6),
\ee
and
\be
0 \rightarrow 
\bar q^{h_1}(p_1)\,
Q^{h_2}(p_2)\,
\bar Q^{h_3}(p_3)\,
q^{h_4}(p_4)\,
\bar \ell^+(p_5)\,
\ell^-(p_6)\, .
\ee
More specifically, we consider the finite remainders defined in Eq.~(2.29) of Ref. \cite{DeLaurentis:2025dxw},
\be
R_\kappa^{(l)[j]}(1^+,2^{h_2},3^{h_3},4^-)
= \sum_i r_{\kappa,i}^{(l)[j] h_2 h_3}\, G_i \,,
\ee
where $\kappa=g$ or $q$ labels the channel, $0\le l\le 2$ the loop
order, and $0\leq j\leq l$ the power of $N_f/N_c$ associated with a given
remainder. The coefficients $r_{\kappa,i}$ are rational functions
of the external kinematics; therefore, their  analytic continuation is trivial, for any choice of momenta of external particles. The functions $G_i$ are special transcendental functions — the one-mass pentagon functions \cite{Chicherin:2021dyp} — originally given in the channel with two massless QCD partons in the initial state. Recently, their analytic continuation for a massive leg from the final to the initial state was presented in Ref.~\cite{Chen:2026jxf}.

Although the helicity remainders do retain the full information about all three physical polarization states, $\varepsilon^+, \varepsilon^-$ and $\varepsilon^L$, recovering the $\varepsilon^L$ remainder from the six point result is not necessarily straightforward (while $\varepsilon^-$ component is easily obtained by reversing the decay current via $5\leftrightarrow 6$). After multiplying through by $s_{56}$, the following map holds
\be\label{eq:six-point-to-five-point-remainder-map}
s_{56} \, R_\kappa^{(l)[j]}(1, 2, 3, 4, 5, 6) \rightarrow R_\kappa^{(l)[j]}(1, 2, 3, 4, \boldsymbol 5)^{IJ},
\ee
where massless legs $5, 6$ are interpreted as the $I=1, 2$ components of $\bold{5}$ as per  Eq.~\eqref{eq:massless-to-massive-spinors}. Crucially, we need to make manifest that each $R_\kappa^{IJ}$ can be written with exactly one power of $[\bold 5|^I$ and $|\bold 5\rangle^J$. We encounter two issues that need to be resolved before this can be made explicit, enabling the recovery of the longitudinal component.

First, some of the spinor-helicity rational functions for the $\bar qgg q \ell\bar\ell$ and $\bar qQ\bar Q q \ell\bar\ell$ remainders have accidental\footnote{We call these symmetries accidental because they are either broken by QCD corrections or are not even symmetries of full tree amplitudes.} symmetries where a quark pair can be swapped with the lepton pair. Such symmetries were used in Ref.~\cite{DeLaurentis:2025dxw}, but now they prevent us from expressing the amplitude with a manifest $[ \boldsymbol 5^I \,|\, \gamma^\mu \,|\, \boldsymbol 5^J \rangle$ current. Therefore, we first need to re-express 
the coefficients using a basis without such symmetries. For instance, one of the NMHV gluon rational basis function reads
\be
\begin{gathered}
r_{g, i} = 
\Bigg[ \frac{[13]⟨24⟩[25]⟨26⟩}{⟨2|1+4|3]^3} \Bigg] - \Bigg[ \, \Bigg]_{123456\rightarrow \overline{432165}}
- \Bigg[ \, \Bigg]_{123456\rightarrow 523614} + \Bigg[ \, \Bigg]_{123456\rightarrow \overline{632541}} \\[2mm] - \frac{3[13]⟨24⟩⟨26⟩[35](s_{124}-s_{134})}{⟨2|1+4|3]^4} \, ,
\end{gathered}
\ee
where the overline on the permutation denotes parity conjugation (swap of angle and square brackets). This function must first be written as,
\be
\begin{gathered}
r_{g, i} = 
\Bigg[ \frac{[13]⟨24⟩[25]⟨26⟩}{⟨2|1+4|3]^3} + \frac{[53]⟨26⟩[21]⟨24⟩}{⟨2|1+4|3]^3} \Bigg] - \Bigg[ \, \Bigg]_{123456\rightarrow \overline{432165}}\\[2mm] - \frac{3[13]⟨24⟩⟨26⟩[35](s_{124}-s_{134})}{⟨2|1+4|3]^4} \, ,
\end{gathered}
\ee
and, once this is done, we can multiply it  by $s_{56}=s_{\boldsymbol 5}$, which is trivial in this case since no $\langle 56\rangle$ or $[56]$ pole is present, and then map $[5| \rightarrow [\boldsymbol 5|^I$ and $|6\rangle \rightarrow |\boldsymbol 5\rangle^J$.

Second, while it should be possible to extract the current  $[ \boldsymbol 5^I \,|\, \gamma^\mu \,|\, \boldsymbol 5^J \rangle$ from each of the rational functions, the $r_{g/q, i}$ are given in a partial fraction decomposition and this property is not always satisfied for each of the terms individually. For instance, some numerators contain differences of Mandelstam variables of the form $(s_{145}-s_{146}) = \trace{(1+4|5-6)}$, where we have made explicit  that this quantity depends on the difference of the momenta $5$ and $6$, while only their sum is admissible if we want to uncover the spin covariance. To solve this issue, we need to refit all rational functions with a manifestly covariant ansatz, imposing degree bounds of 1 for the states $[\boldsymbol q^I|$ and $|\, \boldsymbol q^J \rangle$, and causing a slight reshuffling of the numerator monomials across the terms of the partial fraction decomposition where necessary. This refitting  effectively automates the procedure to build the operator $\mathcal{O}$.

We provide massive-spinor helicity expressions for\footnote{Following the map in Eq.~\eqref{eq:six-point-to-five-point-remainder-map} we only remove the $s_{\boldsymbol 5}$ propagator, i.e. an extra factor of $1/(\sqrt{2}m_{\boldsymbol 5})$ for the transverse or $1/(2m_{\boldsymbol 5})$ for the longitudinal polarization is  required to match the ortho-normal definitions introduced in Section \ref{sect:2}.}
\be
R_\kappa^{(l)[j]}(1, 2, 3, 4, \boldsymbol 5)^{IJ} = \begin{pmatrix} R^- & R^L/2 \\ R^L/2  & R^+ \end{pmatrix}, \quad \forall \kappa=\{g, q\}, \; 0\le l\le 2, \; 0\leq j\leq l \, .
\label{Remainders}
\ee
Due to the ambiguity explained between Eqs.~\eqref{eq:ward-identity-six-point} and \eqref{eq:spin-covariant-amplitude}, the off-diagonal entries are not directly equal to $R^L/2$, but they do sum to $R^L$. The symmetrized choice is the one that makes the fulfillment of the Ward identity manifest. Numerical reference 
values for various finite remainders  are provided  
in Appendix~\ref{app:a}.

%%%%%%%%%%%%%%%%%%%%%%%%%%%%%%%%%%%%%%%%%%%%%%%%%%%%%%%%%%%%%%
%                        CONCLUSIONS                         %
%%%%%%%%%%%%%%%%%%%%%%%%%%%%%%%%%%%%%%%%%%%%%%%%%%%%%%%%%%%%%%

\section{Conclusions}
\label{sect:conclusions}

Scattering amplitudes with external electroweak vector bosons are often computed using matrix elements of the vector current $[6|\gamma^\mu|5\rangle$ in place of the vector boson polarization vector. This is motivated by the fact that vector bosons are most readily observed through their decays to (effectively massless) leptons. Indeed, if the four-momenta $p_{5,6}$ in the current are identified with the momenta of a lightlike lepton and anti-lepton pair, subject to the constraint $q=p_5+p_6$, where $q$ is the momentum of the vector boson, physically relevant amplitudes are obtained immediately. 

It has often been thought that this approach has one drawback. Namely, that amplitudes computed in this way cannot be directly used to describe processes where an intermediate vector boson transitions into final states that are more complex than a pair of massless fermions, since amplitudes for the longitudinally polarized massive (or off-shell) vector boson are also required in such cases.

In this paper, we have shown that this assertion is wrong, and that amplitudes corresponding to the \textit{transverse} polarizations retain the full information required to reconstruct the \textit{longitudinal} one. Consequently, amplitudes for all vector-boson polarization states can be obtained from existing spinor-helicity results that use the transition $V^*\to e^+e^-$ current as a proxy for the vector boson polarization, see e.g.  Refs~\cite{Bern_1998,DeLaurentis:2025dxw} for the one-loop full-color and two-loop leading-color amplitudes. In fact, when starting from suitably organized expressions, the procedure is straightforward and amounts to simple replacement rules for the electron and positron spinors.

Following this procedure, we constructed the helicity amplitudes with a longitudinally polarized vector boson for the processes  $V\to 3j$ and $V\to 4j$, as well as processes related to those by crossing symmetry. We verified the correctness of the results by comparing the one-loop amplitudes against an independent diagrammatic calculation, and by matching the two-loop leading-color remainders against the three form factors derived  in Ref.~\cite{Abreu:2021asb}. The results for the one-loop amplitudes in the FDH scheme, together with the one- and two-loop finite remainders in the tHV scheme, with Catani operators used for infrared subtraction, are provided in the ancillary files accompanying this work. To the best of our knowledge, this provides the first example of two-loop five-point amplitudes written explicitly in the massive spinor-helicity formalism.

The procedure to recover the longitudinal amplitude from the transverse one can be understood in terms of the little-group covariance of the amplitude with respect to the massive vector-boson momentum. In this formulation, the amplitude transforms as a tensor with two open $\mathrm{SU}(2)$ little-group indices. The longitudinal and transverse amplitudes correspond to different linear combinations of the entries of the resulting two-by-two amplitude matrix. Knowledge of any single component is therefore sufficient to reconstruct the full covariant amplitude.
Finally, we stress that this result relies only on the analytic properties of helicity amplitudes and is therefore independent of the loop order and of the details of the process under consideration.

%-----------------------------------------------------
%                     ACKNOWLEDGMENTS
%-----------------------------------------------------

\acknowledgments

M.T. is grateful to Ming-Ming Long for useful conversations.
The research of K.M. is supported by the Deutsche Forschungsgemeinschaft (DFG, German Research Foundation) under grant no.\ 396021762 - TRR 257. 
The research of M.T. is supported by a grant from Deutscher Akademischer Austauschdienst (DAAD). G.D.L.'s work is supported in part by the U.K.\ Royal Society through Grant URF\textbackslash R1\textbackslash 20109.

%-----------------------------------------------------
%                     APPENDICES
%-----------------------------------------------------
\newpage 

\appendix
%%%%%%%%%%%%%%%%%%%%%%%%%%%%%%%%%%%%%%%%%%%%%%%%%%%%%%%%%%%%%%
%                      APPENDIX A                            %
%%%%%%%%%%%%%%%%%%%%%%%%%%%%%%%%%%%%%%%%%%%%%%%%%%%%%%%%%%%%%%

\section{Reference point for  numerical evaluation}
\label{app:a}

For illustration purposes, in this appendix we present the numerical evaluation of the one- and two-loop amplitudes for the reference phase-space point of Eq.~(C.1) in Ref.~\cite{Abreu:2021asb},
\begin{align}
p_i^{\,\mu} =
\begin{pmatrix}
  - 1.54784335 &  \m0.17060632 &  \m1.24932185 &  - 0.89772347 \\
  - 2.02815803 &  \m0.20134469 &  \m1.47441384 &  \m1.37803815 \\
 \m1.71661712 &   - 0.67006338 &  - 1.23465572 &  - 0.98661778 \\
 \m0.82499780 &   - 0.23369413 &  - 0.78062204 &  \m0.12898705 \\
 \m0.94057762 &   \m0.58200384 &  - 0.62974324 &  \m0.38649871 \\
 \m0.09380884 &   - 0.05019734 &  - 0.07871470 &  - 0.00918266
\end{pmatrix} \, 
\label{PhaseSpacePoint},
\end{align}
where legs 1 and 2 are taken to be the incoming ones.

In Tables~\ref{table_Vqqbgg}--\ref{table_VqqbQQb} we report values of all primitive amplitudes obtained from Ref.~\cite{Bern_1998} for a longitudinally-polarized vector boson for the above phase-space point.

On the other hand, in Tables~\ref{tab:finite_gluonspp_pp}--\ref{tab:finite_quarkspm_pm} we provide values for the (un-normalized) remainders for all components of the amplitudes defined in Eq.~(\ref{Remainders}). The ratio of the loop components to the corresponding tree, i.e.~the $(l,j)=(0,0)$ entry, in the first column reproduces the values in Table~5 of Ref.~\cite{Abreu:2021asb} and the leading-color amplitudes in Tables~\ref{table_Vqqbgg}--\ref{table_VqqbQQb}. Target values are generated using the ancillary files of Ref.~\cite{Abreu:2021asb} and modifying the lepton current, as explained above. These are then matched to the massive-spinor helicity expressions we construct.

Analytic results are available on 
\href{https://github.com/GDeLaurentis/antares-results}{GitHub} 
and archived on Zenodo \cite{giuseppe_de_laurentis_2026_18895380}. 
The repository includes \texttt{pytest} scripts demonstrating how to evaluate and assemble the finite remainder, reproducing the values reported in this work. 
These tests are automatically executed in a continuous-integration workflow 
(\href{https://github.com/GDeLaurentis/antares-results/actions/workflows/ci_test.yml}{CI}). 
Human-readable expressions are available in the 
\href{https://gdelaurentis.github.io/antares-results/antares_results/Vjj/Vjj.html}{online documentation}.

As an illustration, after installing the package via \texttt{pip install antares-results}, the analytic expressions can be imported and evaluated directly in Python.\footnote{To export to Mathematica (\texttt{S@M} format \cite{Maitre_2007}), use the \texttt{Terms.toSaM} function.} For example,
\begin{verbatim}
from antares_results.Vjj.momenta import oPsAllUp
from antares_results.Vjj.qggqV.nmhv import lTerms

lTerms(oPsAllUp)
\end{verbatim}
This evaluates at the phase-space point of Eq.~(\ref{PhaseSpacePoint}) the rational coefficients entering the finite remainders 
$R_g^{(l)[j]}(\bar q_1^+, g_2^\pm, g_3^\mp, q_4^-, V_{\boldsymbol{5}})$, returning them with the two $\mathrm{SU}(2)$ spin indices in the upper position.

\begin{table}[ht]
    \centering
    \resizebox{\textwidth}{!}{%
    \begin{tabular}{cccc}
        \hline
        \multicolumn{4}{c}{\textbf{One-loop amplitudes for $\boldsymbol{V \to q\bar{q}gg}$}} \\
        \hline
            \textbf{Primitive Amplitude}
            & $\boldsymbol{\epsilon^{-2}}$
            & $\boldsymbol{\epsilon^{-1}}$
            & $\boldsymbol{\epsilon^{0}}$
            \\ \hline \vspace{0.5mm} 
            %%%%%%%%%%%%%%%%%%%%%%%%%%%%%%%%%%%%%%%%%%%%%%%%%%%%%%%%%%%%%
            % Numbers
            %%%%%%%%%%%%%%%%%%%%%%%%%%%%%%%%%%%%%%%%%%%%%%%%%%%%%%%%%%%%%
            $A_{6}^{\text{s}}(1_{q}^{+},2^{+},3^{+},4_{\bar{q}}^{-})$    
            & 0
            & 0
            & $-5.5476 + 4.6069 \, \rm{i}$
            \\
            %%%%%%%%%%%%%%%%%%%%%%%%%%%%%%%%%%%%%%%%%%%%%%%%%%%%%%%%%%%%%
            $A_{6}^{\text{s}}(1_{q}^{+},2^{+},3^{-},4_{\bar{q}}^{-})$      
            & 0
            & 0
            & 0
            \\
            %%%%%%%%%%%%%%%%%%%%%%%%%%%%%%%%%%%%%%%%%%%%%%%%%%%%%%%%%%%%%
            $A_{6}^{\text{s}}(1_{q}^{+},2^{-},3^{+},4_{\bar{q}}^{-})$ 
            & 0
            & 0
            & 0
            \\[0.5mm] \hline
            %%%%%%%%%%%%%%%%%%%%%%%%%%%%%%%%%%%%%%%%%%%%%%%%%%%%%%%%%%%%%
            $A_{6}(1_{q}^{+},2^{+},3^{+},4_{\bar{q}}^{-})$    
            & $-3.0000$
            & $-1.1562 - 6.2832 \, \rm{i}$
            & $0.9887 - 0.7294 \, \rm{i}$
            \\ 
            %%%%%%%%%%%%%%%%%%%%%%%%%%%%%%%%%%%%%%%%%%%%%%%%%%%%%%%%%%%%%
            $A_{6}(1_{q}^{+},2^{+},3^{-},4_{\bar{q}}^{-})$      
            & $-3.0000$
            & $-1.1562 - 6.2832 \, \rm{i}$
            & $7.5920 - 8.3969 \, \rm{i}$
            \\ 
            %%%%%%%%%%%%%%%%%%%%%%%%%%%%%%%%%%%%%%%%%%%%%%%%%%%%%%%%%%%%%
            $A_{6}(1_{q}^{+},2^{-},3^{+},4_{\bar{q}}^{-})$ 
            & $-3.0000$
            & $-1.1562 - 6.2832 \, \rm{i}$
            & $8.6675 - 3.2040 \, \rm{i}$
            \\[0.5mm] \hline
            %%%%%%%%%%%%%%%%%%%%%%%%%%%%%%%%%%%%%%%%%%%%%%%%%%%%%%%%%%%%%
            $A_{6}(1_{q}^{+},2^{+},3_{\bar{q}}^{-},4^{+})$    
            & $-2.0000$
            & $-0.9892 - 3.1416 \, \rm{i}$
            & $\m 0.2755 - 8.9182 \, \rm{i}$
            \\ 
            %%%%%%%%%%%%%%%%%%%%%%%%%%%%%%%%%%%%%%%%%%%%%%%%%%%%%%%%%%%%%
            $A_{6}(1_{q}^{+},2^{+},3_{\bar{q}}^{-},4^{-})$
            & $-2.0000$
            & $-0.9892 - 3.1416 \, \rm{i}$
            & $-5.7728 - 22.8208 \, \rm{i}$
           \\[0.5mm] \hline
            %%%%%%%%%%%%%%%%%%%%%%%%%%%%%%%%%%%%%%%%%%%%%%%%%%%%%%%%%%%%%
            $A_{6}(1_{q}^{+},2_{\bar{q}}^{-},3^{+},4^{+})$    
            & $-1.0000$
            & $\m 0.1094 - 3.1416 \, \rm{i}$
            & $\m 0.7823 - 6.9762 \, \rm{i}$
            \\ 
            %%%%%%%%%%%%%%%%%%%%%%%%%%%%%%%%%%%%%%%%%%%%%%%%%%%%%%%%%%%%%
            $A_{6}(1_{q}^{+},2_{\bar{q}}^{-},3^{+},4^{-})$     
            & $-1.0000$
            & $\m 0.1094 - 3.1416 \, \rm{i}$
            & $-3.8985 - 15.7853 \, \rm{i}$
            \\ 
            %%%%%%%%%%%%%%%%%%%%%%%%%%%%%%%%%%%%%%%%%%%%%%%%%%%%%%%%%%%%%
            $A_{6}(1_{q}^{+},2_{\bar{q}}^{-},3^{-},4^{+})$     
            & $-1.0000$
            & $\m 0.1094 - 3.1416 \, \rm{i}$
            & $-0.9315 - 9.2910 \, \rm{i}$
            \\ \hline
            $A_{6}^{\text{v}}(1_{q}^{+},2_{\bar{q}}^{-},3^{+},4^{+})$    
            & 0
            & 0
            & $0.0089 + 0.0146 \, \rm{i}$
            \\ 
            %%%%%%%%%%%%%%%%%%%%%%%%%%%%%%%%%%%%%%%%%%%%%%%%%%%%%%%%%%%%%
            $A_{6}^{\text{v}}(1_{q}^{+},2_{\bar{q}}^{-},3^{+},4^{-})$ 
            & 0
            & 0
            & $0.1269 - 0.0247  \, \rm{i}$
            \\[0.5mm] \hline
    \end{tabular}
    }
    \caption{Ratios of one-loop to tree amplitudes for the ``longitudinally-polarized'' vector boson decaying to $q\bar{q}gg$ evaluated at the phase-space point of Eq.~(\ref{PhaseSpacePoint}). We follow the same convention for the definition of   helicity amplitudes as in Ref.~\cite{Bern_1998}, except of course for the polarization vector of the vector boson which is taken to be $\epsilon_q^{(L),\mu}$, see Eq.~(\ref{eq:massless-spinor-longitudinal-polarization}). We note that in the notation of Ref.~\cite{Bern_1998} the functions $A_{6}(1_{q},2,3,4_{\bar{q}}) \, , \, A_{6}(1_{q},2,3_{\bar{q}},4)$ and $A_{6}(1_{q},2_{\bar{q}},3,4)$ correspond to \emph{different} primitive amplitudes and not to the same amplitude for different orderings of the QCD partons. Spinor products were evaluated with the package \texttt{S@M}~\cite{Maitre_2007}.}
    \label{table_Vqqbgg}
\end{table}
\begin{table}[ht]
    \centering
    \resizebox{\textwidth}{!}{%
    \begin{tabular}{cccc}
        \hline
        \multicolumn{4}{c}{\textbf{One-loop amplitudes for $\boldsymbol{V \to q\bar{q}Q\bar{Q}}$}} \\
        \hline
            \textbf{Primitive Amplitude}
            & $\boldsymbol{\epsilon^{-2}}$
            & $\boldsymbol{\epsilon^{-1}}$
            & $\boldsymbol{\epsilon^{0}}$
            \\ \hline \vspace{0.5mm} 
            %%%%%%%%%%%%%%%%%%%%%%%%%%%%%%%%%%%%%%%%%%%%%%%%%%%%%%%%%%%%%
            % Numbers
            %%%%%%%%%%%%%%%%%%%%%%%%%%%%%%%%%%%%%%%%%%%%%%%%%%%%%%%%%%%%%
            $A_{6}^{\text{f}}(1_{q}^{+},2_{\bar{Q}}^{+},3_{Q}^{-},4_{\bar{q}}^{+})$    
            & 0
            & $0.6667$
            & $1.8435$
            \\ 
            %%%%%%%%%%%%%%%%%%%%%%%%%%%%%%%%%%%%%%%%%%%%%%%%%%%%%%%%%%%%%
            $A_{6}^{\text{f}}(1_{q}^{+},2_{\bar{Q}}^{-},3_{Q}^{+},4_{\bar{q}}^{+})$     
            & 0
            & $0.6667$
            & $1.8435$
            \\[0.5mm] \hline
            %%%%%%%%%%%%%%%%%%%%%%%%%%%%%%%%%%%%%%%%%%%%%%%%%%%%%%%%%%%%%
            $A_{6}(1_{q}^{+},2_{\bar{Q}}^{+},3_{Q}^{-},4_{\bar{q}}^{+})$    
            & $-2.0000$
            & $\m 2.1091 - 6.2832 \, \rm{i}$
            & $15.9512 - 4.2722 \, \rm{i}$
            \\
            %%%%%%%%%%%%%%%%%%%%%%%%%%%%%%%%%%%%%%%%%%%%%%%%%%%%%%%%%%%%%
            $A_{6}(1_{q}^{+},2_{\bar{Q}}^{-},3_{Q}^{+},4_{\bar{q}}^{+})$     
            & $-2.0000$
            & $\m 2.1091 - 6.2832 \, \rm{i}$
            & $15.5544 - 7.2669 \, \rm{i}$
           \\[0.5mm] \hline
            %%%%%%%%%%%%%%%%%%%%%%%%%%%%%%%%%%%%%%%%%%%%%%%%%%%%%%%%%%%%%
            $A_{6}^{\text{sl}}(1_{q}^{+},2_{\bar{Q}}^{-},3_{Q}^{+},4_{\bar{q}}^{+})$     
            & $-2.0000$
            & $-1.5576 - 6.2832 \, \rm{i}$
            & $\m 6.8870 - 2.5255 \, \rm{i}$
            \\[0.5mm] \hline
    \end{tabular}
    }
    \caption{Ratios of one-loop to tree amplitudes for the ``longitudinally-polarized'' vector boson decaying to $q\bar{q}Q\Bar{Q}$ evaluated at the phase-space point of Eq.~(\ref{PhaseSpacePoint}). We follow the same convention for the definition of   helicity amplitudes as in Ref.~\cite{Bern_1998}, except of course for the polarization vector of the vector boson which is taken to be $\epsilon_q^{(L),\mu}$, see Eq.~(\ref{eq:massless-spinor-longitudinal-polarization}).}
    \label{table_VqqbQQb}
\end{table}

\begin{table}[ht]
\centering
\resizebox{\textwidth}{!}{%
\begin{tabular}{cccc}
\hline
\multicolumn{4}{c}{$\boldsymbol{R_g^{(l)[j]}(\bar q_1^+, g_2^+, g_3^+, q_4^-, V_{\boldsymbol {5}})}$} \\[0.1mm]
\hline
{$(l, j)$} & ${R^{+}}$ & ${R^{-}}$ & ${R^{L}}$ \\
\hline
$(0,0)$ & $0.0236 + 0.0045\,\mathrm{i}$ & $1.4518 + 0.7864\,\mathrm{i}$ & $-0.3756 - 0.1334\,\mathrm{i}$ \\
$(1,0)$ & $2.5170 + 0.0688\,\mathrm{i}$ & $-7.9473 + 5.8586\,\mathrm{i}$ & $3.6470 - 0.9089\,\mathrm{i}$ \\
$(1,1)$ & $-1.9910 + 0.1432\,\mathrm{i}$ & $7.3012 - 0.7559\,\mathrm{i}$ & $-2.7909 + 1.4004\,\mathrm{i}$ \\
$(2,0)$ & $66.8648 - 13.0502\,\mathrm{i}$ & $-63.7138 + 142.0155\,\mathrm{i}$ & $69.7945 - 15.6408\,\mathrm{i}$ \\
$(2,1)$ & $-57.4965 + 16.2763\,\mathrm{i}$ & $158.0937 - 29.3544\,\mathrm{i}$ & $-76.7995 + 26.2648\,\mathrm{i}$ \\
$(2,2)$ & $4.0727 + 1.7690\,\mathrm{i}$ & $-14.3374 - 8.3390\,\mathrm{i}$ & $6.8454 + 0.7164\,\mathrm{i}$ \\
\hline
\end{tabular}
}
\caption{Finite remainders for the three $V^{\{+,-,L\}}$ polarization states for $\bar q_1^+, g_2^+, g_3^+, q_4^-, V_{\boldsymbol {5}}$.}
\label{tab:finite_gluonspp_pp}
\end{table}
\begin{table}[ht]
\centering
\resizebox{\textwidth}{!}{%
\begin{tabular}{cccc}
\hline
\multicolumn{4}{c}{$\boldsymbol{R_g^{(l)[j]}(\bar q_1^+, g_2^-, g_3^+, q_4^-, V_{\boldsymbol {5}})}$} \\[0.1mm]
\hline
{$(l, j)$} & ${R^{+}}$ & ${R^{-}}$ & ${R^{L}}$ \\
\hline
$(0,0)$ & $-8.8426 + 3.4839\,\mathrm{i}$ & $13.7981 - 5.5470\,\mathrm{i}$ & $-6.5913 + 7.4813\,\mathrm{i}$ \\
$(1,0)$ & $34.8818 + 47.4393\,\mathrm{i}$ & $-63.3181 - 15.2968\,\mathrm{i}$ & $26.6477 + 6.8022\,\mathrm{i}$ \\
$(1,1)$ & $4.7607 + 8.8217\,\mathrm{i}$ & $-7.5447 - 13.7515\,\mathrm{i}$ & $8.6636 + 5.9612\,\mathrm{i}$ \\
$(2,0)$ & $-84.6750 + 1060.7588\,\mathrm{i}$ & $-20.6528 - 1489.2675\,\mathrm{i}$ & $291.1200 + 643.6633\,\mathrm{i}$ \\
$(2,1)$ & $42.5230 - 288.7742\,\mathrm{i}$ & $80.9694 + 358.5608\,\mathrm{i}$ & $-82.9875 - 151.6379\,\mathrm{i}$ \\
$(2,2)$ & $6.3927 - 10.6071\,\mathrm{i}$ & $-9.8875 + 16.6661\,\mathrm{i}$ & $0.8976 - 12.9615\,\mathrm{i}$ \\
\hline
\end{tabular}
}
\caption{Finite remainders for the three $V^{\{+,-,L\}}$ polarization states for $\bar q_1^+, g_2^-, g_3^+, q_4^-, V_{\boldsymbol {5}}$.}
\label{tab:finite_gluonsmp_mp}
\end{table}
\begin{table}[ht]
\centering
\resizebox{\textwidth}{!}{%
\begin{tabular}{cccc}
\hline
\multicolumn{4}{c}{$\boldsymbol{R_g^{(l)[j]}(\bar q_1^+, g_2^+, g_3^-, q_4^-, V_{\boldsymbol {5}})}$} \\[0.1mm]
\hline
{$(l, j)$} & ${R^{+}}$ & ${R^{-}}$ & ${R^{L}}$ \\
\hline
$(0,0)$ & $3.7229 + 1.7267\,\mathrm{i}$ & $-24.5982 - 17.0571\,\mathrm{i}$ & $6.9476 + 2.2245\,\mathrm{i}$ \\
$(1,0)$ & $9.4393 - 20.0244\,\mathrm{i}$ & $57.2350 + 96.5030\,\mathrm{i}$ & $-7.3337 + 18.6134\,\mathrm{i}$ \\
$(1,1)$ & $1.3398 - 4.1159\,\mathrm{i}$ & $-14.7675 + 27.9051\,\mathrm{i}$ & $1.4554 - 7.5553\,\mathrm{i}$ \\
$(2,0)$ & $464.4059 - 252.9157\,\mathrm{i}$ & $-1920.6393 + 1177.2971\,\mathrm{i}$ & $161.1832 - 60.1954\,\mathrm{i}$ \\
$(2,1)$ & $-114.3829 + 48.6404\,\mathrm{i}$ & $366.5085 - 383.4010\,\mathrm{i}$ & $14.4674 + 3.3703\,\mathrm{i}$ \\
$(2,2)$ & $-5.2201 + 1.1609\,\mathrm{i}$ & $38.9628 - 1.8247\,\mathrm{i}$ & $-8.9514 + 3.1990\,\mathrm{i}$ \\
\hline
\end{tabular}
}
\caption{Finite remainders for the three $V^{\{+,-,L\}}$ polarization states for $\bar q_1^+, g_2^+, g_3^-, q_4^-, V_{\boldsymbol {5}}$.}
\label{tab:finite_gluonspm_pm}
\end{table}
\begin{table}[ht]
\centering
\resizebox{\textwidth}{!}{%
\begin{tabular}{cccc}
\hline
\multicolumn{4}{c}{$\boldsymbol{R_q^{(l)[j]}(\bar q_1^+, Q_2^-,  \bar Q_3^+, q_4^-, V_{\boldsymbol {5}})}$} \\[0.1mm]
\hline
{$(l, j)$} & ${R^{+}}$ & ${R^{-}}$ & ${R^{L}}$ \\
\hline
$(0,0)$ & $6.3175 + 4.1683\,\mathrm{i}$ & $-13.4644 - 10.9564\,\mathrm{i}$ & $8.9602 + 1.8117\,\mathrm{i}$ \\
$(1,0)$ & $53.7538 - 29.5378\,\mathrm{i}$ & $-53.6393 + 28.8551\,\mathrm{i}$ & $41.6072 - 13.7234\,\mathrm{i}$ \\
$(1,1)$ & $-11.6464 - 7.6843\,\mathrm{i}$ & $24.8218 + 20.1983\,\mathrm{i}$ & $-16.5184 - 3.3399\,\mathrm{i}$ \\
$(2,0)$ & $1277.3178 - 431.2926\,\mathrm{i}$ & $-2158.2100 + 463.4091\,\mathrm{i}$ & $936.4298 - 399.3205\,\mathrm{i}$ \\
$(2,1)$ & $-460.4530 + 70.8618\,\mathrm{i}$ & $733.6914 - 6.0067\,\mathrm{i}$ & $-385.0000 + 119.1601\,\mathrm{i}$ \\
$(2,2)$ & $21.4703 + 14.1662\,\mathrm{i}$ & $-45.7595 - 37.2360\,\mathrm{i}$ & $30.4520 + 6.1572\,\mathrm{i}$ \\
\hline
\end{tabular}
}
\caption{Finite remainders for the three $V^{\{+,-,L\}}$ polarization states for $\bar q_1^+, Q_2^-,  \bar Q_3^+, q_4^-, V_{\boldsymbol {5}}$.}
\label{tab:finite_quarksmp_mp}
\end{table}
\begin{table}[ht]
\centering
\resizebox{\textwidth}{!}{%
\begin{tabular}{cccc}
\hline
\multicolumn{4}{c}{$\boldsymbol{R_q^{(l)[j]}(\bar q_1^+, Q_2^+,  \bar Q_3^-, q_4^-, V_{\boldsymbol {5}})}$} \\[0.1mm]
\hline
{$(l, j)$} & ${R^{+}}$ & ${R^{-}}$ & ${R^{L}}$ \\
\hline
$(0,0)$ & $3.7637 - 2.8566\,\mathrm{i}$ & $-22.1553 + 7.2210\,\mathrm{i}$ & $6.0085 - 5.8687\,\mathrm{i}$ \\
$(1,0)$ & $-2.9488 - 43.2704\,\mathrm{i}$ & $15.0449 + 63.3248\,\mathrm{i}$ & $31.0476 - 23.0314\,\mathrm{i}$ \\
$(1,1)$ & $-6.9384 + 5.2661\,\mathrm{i}$ & $40.8437 - 13.3120\,\mathrm{i}$ & $-11.0768 + 10.8191\,\mathrm{i}$ \\
$(2,0)$ & $100.5951 - 1013.7803\,\mathrm{i}$ & $-572.7447 + 2742.6868\,\mathrm{i}$ & $413.4240 - 606.2951\,\mathrm{i}$ \\
$(2,1)$ & $-72.4095 + 310.0332\,\mathrm{i}$ & $444.4454 - 822.4600\,\mathrm{i}$ & $-166.0346 + 266.8854\,\mathrm{i}$ \\
$(2,2)$ & $12.7911 - 9.7082\,\mathrm{i}$ & $-75.2962 + 24.5409\,\mathrm{i}$ & $20.4202 - 19.9451\,\mathrm{i}$ \\
\hline
\end{tabular}
}
\caption{Finite remainders for the three $V^{\{+,-,L\}}$ polarization states for $\bar q_1^+, Q_2^+,  \bar Q_3^-, q_4^-, V_{\boldsymbol {5}}$.}
\label{tab:finite_quarkspm_pm}
\end{table}

\clearpage

\bibliographystyle{JHEP}
\bibliography{biblio.bib}

@article{Pellen:2021vpi,
    author = "Pellen, Mathieu and Poncelet, Rene and Popescu, Andrei",
    title = "{Polarised W+j production at the LHC: a study at NNLO QCD accuracy}",
    eprint = "2109.14336",
    archivePrefix = "arXiv",
    primaryClass = "hep-ph",
    reportNumber = "CAVENDISH-HEP-21/13, FR-PHENO-2021-11",
    doi = "10.1007/JHEP02(2022)160",
    journal = "JHEP",
    volume = "02",
    pages = "160",
    year = "2022"
}

@article{Gauld:2021ule,
    author = "Gauld, R. and Gehrmann-De Ridder, A. and Glover, E. W. N. and Huss, A. and Majer, I.",
    title = "{VH + jet production in hadron-hadron collisions up to order $ {\alpha}_{\mathrm{s}}^3 $ in perturbative QCD}",
    eprint = "2110.12992",
    archivePrefix = "arXiv",
    primaryClass = "hep-ph",
    reportNumber = "NIKHEF 2021-026, BONN-TH-2021-09, IPPP/21/26, ZU-TH 49/21,
  CERN-TH-2021-159",
    doi = "10.1007/JHEP03(2022)008",
    journal = "JHEP",
    volume = "03",
    pages = "008",
    year = "2022"
}

@article{Ossola:2007ax,
    author = "Ossola, Giovanni and Papadopoulos, Costas G. and Pittau, Roberto",
    title = "{CutTools: A Program implementing the OPP reduction method to compute one-loop amplitudes}",
    eprint = "0711.3596",
    archivePrefix = "arXiv",
    primaryClass = "hep-ph",
    doi = "10.1088/1126-6708/2008/03/042",
    journal = "JHEP",
    volume = "03",
    pages = "042",
    year = "2008"
}

@article{Ossola:2006us,
    author = "Ossola, Giovanni and Papadopoulos, Costas G. and Pittau, Roberto",
    title = "{Reducing full one-loop amplitudes to scalar integrals at the integrand level}",
    eprint = "hep-ph/0609007",
    archivePrefix = "arXiv",
    doi = "10.1016/j.nuclphysb.2006.11.012",
    journal = "Nucl. Phys. B",
    volume = "763",
    pages = "147--169",
    year = "2007"
}

@article{Berger:2008sj,
    author = "Berger, C. F. and Bern, Z. and Dixon, L. J. and Febres Cordero, F. and Forde, D. and Ita, H. and Kosower, D. A. and Maitre, D.",
    title = "{An Automated Implementation of On-Shell Methods for One-Loop Amplitudes}",
    eprint = "0803.4180",
    archivePrefix = "arXiv",
    primaryClass = "hep-ph",
    reportNumber = "SLAC-PUB-13161, UCLA-08-TEP-10, MIT-CTP-3937, NSF-KITP-08-48, SACLAY-IPHT-T08-054",
    doi = "10.1103/PhysRevD.78.036003",
    journal = "Phys. Rev. D",
    volume = "78",
    pages = "036003",
    year = "2008"
}

@article{Denner:2016kdg,
    author = "Denner, Ansgar and Dittmaier, Stefan and Hofer, Lars",
    title = "{Collier: a fortran-based Complex One-Loop LIbrary in Extended Regularizations}",
    eprint = "1604.06792",
    archivePrefix = "arXiv",
    primaryClass = "hep-ph",
    reportNumber = "FR-PHENO-2016-003, ICCUB-16-016",
    doi = "10.1016/j.cpc.2016.10.013",
    journal = "Comput. Phys. Commun.",
    volume = "212",
    pages = "220--238",
    year = "2017"
}

@article{Buccioni:2019sur,
    author = {Buccioni, Federico and Lang, Jean-Nicolas and Lindert, Jonas M. and Maierh{\"o}fer, Philipp and Pozzorini, Stefano and Zhang, Hantian and Zoller, Max F.},
    title = "{OpenLoops 2}",
    eprint = "1907.13071",
    archivePrefix = "arXiv",
    primaryClass = "hep-ph",
    reportNumber = "IPPP/19/62, FR-PHENO-2019-12, PSI-PR-19-15, ZU-TH 37/19",
    doi = "10.1140/epjc/s10052-019-7306-2",
    journal = "Eur. Phys. J. C",
    volume = "79",
    number = "10",
    pages = "866",
    year = "2019"
}

@article{Bern_1998,
   title={One-loop amplitudes for e+e-- to four partons},
   volume={513},
   ISSN={0550-3213},
   url={http://dx.doi.org/10.1016/S0550-3213(97)00703-7},
   DOI={10.1016/s0550-3213(97)00703-7},
   number={1–2},
   journal={Nuclear Physics B},
   publisher={Elsevier BV},
   author={Bern, Zvi and Dixon, Lance and Kosower, David A.},
   year={1998},
   month=mar, pages={3–86} }

@article{Garland_2002,
   title={Two-loop QCD helicity amplitudes for e+e--→3jets},
   volume={642},
   ISSN={0550-3213},
   url={http://dx.doi.org/10.1016/S0550-3213(02)00627-2},
   DOI={10.1016/s0550-3213(02)00627-2},
   number={1–2},
   journal={Nuclear Physics B},
   publisher={Elsevier BV},
   author={Garland, L.W. and Gehrmann, T. and Glover, E.W.N. and Koukoutsakis, A. and Remiddi, E.},
   year={2002},
   month=oct, pages={227–262} }

@article{Gehrmann_2023,
   title={Two-loop QCD corrections to the V → $q\bar{q}g$ helicity amplitudes with axial-vector couplings},
   volume={2023},
   ISSN={1029-8479},
   url={http://dx.doi.org/10.1007/JHEP02(2023)041},
   DOI={10.1007/jhep02(2023)041},
   number={2},
   journal={Journal of High Energy Physics},
   publisher={Springer Science and Business Media LLC},
   author={Gehrmann, Thomas and Peraro, Tiziano and Tancredi, Lorenzo},
   year={2023},
   month=feb }

@article{Bern_1991aq,
    author = "Bern, Zvi and Kosower, David A.",
    title = "{The Computation of Loop Amplitudes in Gauge Theories}",
    reportNumber = "FERMILAB-PUB-91-111-T, PITT-91-05",
    doi = "10.1016/0550-3213(92)90134-W",
    journal = "Nucl. Phys. B",
    volume = "379",
    pages = "451--561",
    year = "1992"
}

@ARTICLE{tHV1973,
       author = {{'t Hooft}, G. and {Veltman}, M.},
        title = "{Regularization and renormalization of gauge fields}",
      journal = {Nuclear Physics B},
         year = 1972,
        month = jul,
       volume = {44},
       number = {1},
        pages = {189-213},
          doi = {10.1016/0550-3213(72)90279-9}
}

@article{Kunszt_1994,
   title={One-loop helicity amplitudes for all 2 → 2 processes in QCD and N = 1 supersymmetric Yang-Mills theory},
   volume={411},
   ISSN={0550-3213},
   url={http://dx.doi.org/10.1016/0550-3213(94)90456-1},
   DOI={10.1016/0550-3213(94)90456-1},
   number={2–3},
   journal={Nuclear Physics B},
   publisher={Elsevier BV},
   author={Kunszt, Zoltan and Signer, Adrian and Trócsányi, Zoltán},
   year={1994},
   month=jan, pages={397–442} }

@article{Peraro_2021,
   title={Tensor decomposition for bosonic and fermionic scattering amplitudes},
   volume={103},
   ISSN={2470-0029},
   url={http://dx.doi.org/10.1103/PhysRevD.103.054042},
   DOI={10.1103/physrevd.103.054042},
   number={5},
   journal={Physical Review D},
   publisher={American Physical Society (APS)},
   author={Peraro, Tiziano and Tancredi, Lorenzo},
   year={2021},
   month=mar }

@article{Peraro_2019,
   title={FiniteFlow: multivariate functional reconstruction using finite fields and dataflow graphs},
   volume={2019},
   ISSN={1029-8479},
   url={http://dx.doi.org/10.1007/JHEP07(2019)031},
   DOI={10.1007/jhep07(2019)031},
   number={7},
   journal={Journal of High Energy Physics},
   publisher={Springer Science and Business Media LLC},
   author={Peraro, Tiziano},
   year={2019},
   month=jul }

@article{Hahn_2001,
   title={Generating Feynman diagrams and amplitudes with FeynArts 3},
   volume={140},
   ISSN={0010-4655},
   url={http://dx.doi.org/10.1016/S0010-4655(01)00290-9},
   DOI={10.1016/s0010-4655(01)00290-9},
   number={3},
   journal={Computer Physics Communications},
   publisher={Elsevier BV},
   author={Hahn, Thomas},
   year={2001},
   month=nov, pages={418–431} }

@article{Mertig_1991,
   title = {Feyn Calc - Computer-algebraic calculation of Feynman amplitudes},
   journal = {Computer Physics Communications},
   volume = {64},
   number = {3},
   pages = {345-359},
   year = {1991},
   issn = {0010-4655},
   doi = {https://doi.org/10.1016/0010-4655(91)90130-D},
   url = {https://www.sciencedirect.com/science/article/pii/001046559190130D},
   author = {R. Mertig and M. Böhm and A. Denner},
}

@article{Shtabovenko_2016,
   title={New developments in FeynCalc 9.0},
   volume={207},
   ISSN={0010-4655},
   url={http://dx.doi.org/10.1016/j.cpc.2016.06.008},
   DOI={10.1016/j.cpc.2016.06.008},
   journal={Computer Physics Communications},
   publisher={Elsevier BV},
   author={Shtabovenko, Vladyslav and Mertig, Rolf and Orellana, Frederik},
   year={2016},
   month=oct, pages={432–444} }

@article{Shtabovenko_2020,
   title={FeynCalc 9.3: New features and improvements},
   volume={256},
   ISSN={0010-4655},
   url={http://dx.doi.org/10.1016/j.cpc.2020.107478},
   DOI={10.1016/j.cpc.2020.107478},
   journal={Computer Physics Communications},
   publisher={Elsevier BV},
   author={Shtabovenko, Vladyslav and Mertig, Rolf and Orellana, Frederik},
   year={2020},
   month=nov, pages={107478} }

@article{Shtabovenko_2025,
   title={FeynCalc 10: Do multiloop integrals dream of computer codes?},
   volume={306},
   ISSN={0010-4655},
   url={http://dx.doi.org/10.1016/j.cpc.2024.109357},
   DOI={10.1016/j.cpc.2024.109357},
   journal={Computer Physics Communications},
   publisher={Elsevier BV},
   author={Shtabovenko, Vladyslav and Mertig, Rolf and Orellana, Frederik},
   year={2025},
   month=jan, pages={109357} }

@misc{Mathematica,
  author = {Wolfram Research{,} Inc.},
  title = {Mathematica, {V}ersion 13.2},
  note = {Champaign, IL, 2022}
}

@article{Kuipers_2013,
   title={FORM version 4.0},
   volume={184},
   ISSN={0010-4655},
   url={http://dx.doi.org/10.1016/j.cpc.2012.12.028},
   DOI={10.1016/j.cpc.2012.12.028},
   number={5},
   journal={Computer Physics Communications},
   publisher={Elsevier BV},
   author={Kuipers, J. and Ueda, T. and Vermaseren, J.A.M. and Vollinga, J.},
   year={2013},
   month=may, pages={1453–1467} }

@article{Maierh_fer_2018,
   title={Kira—A Feynman integral reduction program},
   volume={230},
   ISSN={0010-4655},
   url={http://dx.doi.org/10.1016/j.cpc.2018.04.012},
   DOI={10.1016/j.cpc.2018.04.012},
   journal={Computer Physics Communications},
   publisher={Elsevier BV},
   author={Maierhöfer, P. and Usovitsch, J. and Uwer, P.},
   year={2018},
   month=sep, pages={99–112} }

@article{Klappert_2021,
   title={Integral reduction with Kira 2.0 and finite field methods},
   volume={266},
   ISSN={0010-4655},
   url={http://dx.doi.org/10.1016/j.cpc.2021.108024},
   DOI={10.1016/j.cpc.2021.108024},
   journal={Computer Physics Communications},
   publisher={Elsevier BV},
   author={Klappert, Jonas and Lange, Fabian and Maierhöfer, Philipp and Usovitsch, Johann},
   year={2021},
   month=sep, pages={108024} }

@misc{lange2025,
      title={Kira 3: integral reduction with efficient seeding and optimized equation selection}, 
      author={Fabian Lange and Johann Usovitsch and Zihao Wu},
      year={2025},
      eprint={2505.20197},
      archivePrefix={arXiv},
      primaryClass={hep-ph},
      url={https://arxiv.org/abs/2505.20197}, 
}

@article{Bern_1993,
   title={Dimensionally regulated one-loop integrals},
   volume={302},
   ISSN={0370-2693},
   url={http://dx.doi.org/10.1016/0370-2693(93)90400-C},
   DOI={10.1016/0370-2693(93)90400-c},
   number={2–3},
   journal={Physics Letters B},
   publisher={Elsevier BV},
   author={Bern, Zvi and Dixon, Lance and Kosower, David A.},
   year={1993},
   month=mar, pages={299–308} }

@article{Bern_1994,
   title={Dimensionally-regulated pentagon integrals},
   volume={412},
   ISSN={0550-3213},
   url={http://dx.doi.org/10.1016/0550-3213(94)90398-0},
   DOI={10.1016/0550-3213(94)90398-0},
   number={3},
   journal={Nuclear Physics B},
   publisher={Elsevier BV},
   author={Bern, Zvi and Dixon, Lance and Kosower, David A.},
   year={1994},
   month=jan, pages={751–816} }

@article{Liu_2023,
   title={AMFlow: A Mathematica package for Feynman integrals computation via auxiliary mass flow},
   volume={283},
   ISSN={0010-4655},
   url={http://dx.doi.org/10.1016/j.cpc.2022.108565},
   DOI={10.1016/j.cpc.2022.108565},
   journal={Computer Physics Communications},
   publisher={Elsevier BV},
   author={Liu, Xiao and Ma, Yan-Qing},
   year={2023},
   month=feb, pages={108565} }

@article{Conde:2016izb,
    author = "Conde, Eduardo and Joung, Euihun and Mkrtchyan, Karapet",
    title = "{Spinor-Helicity Three-Point Amplitudes from Local Cubic Interactions}",
    eprint = "1605.07402",
    archivePrefix = "arXiv",
    primaryClass = "hep-th",
    doi = "10.1007/JHEP08(2016)040",
    journal = "JHEP",
    volume = "08",
    pages = "040",
    year = "2016"
}

@article{Conde:2016vxs,
    author = "Conde, Eduardo and Marzolla, Andrea",
    title = "{Lorentz Constraints on Massive Three-Point Amplitudes}",
    eprint = "1601.08113",
    archivePrefix = "arXiv",
    primaryClass = "hep-th",
    doi = "10.1007/JHEP09(2016)041",
    journal = "JHEP",
    volume = "09",
    pages = "041",
    year = "2016"
}

@article{Arkani-Hamed:2017jhn,
    author = "Arkani-Hamed, Nima and Huang, Tzu-Chen and Huang, Yu-tin",
    title = "{Scattering amplitudes for all masses and spins}",
    eprint = "1709.04891",
    archivePrefix = "arXiv",
    primaryClass = "hep-th",
    reportNumber = "NCTS-TH/1714, NCTS-TH-1714",
    doi = "10.1007/JHEP11(2021)070",
    journal = "JHEP",
    volume = "11",
    pages = "070",
    year = "2021"
}

@article{Ochirov:2018uyq,
    author = "Ochirov, Alexander",
    title = "{Helicity amplitudes for QCD with massive quarks}",
    eprint = "1802.06730",
    archivePrefix = "arXiv",
    primaryClass = "hep-ph",
    doi = "10.1007/JHEP04(2018)089",
    journal = "JHEP",
    volume = "04",
    pages = "089",
    year = "2018"
}

@article{Shadmi:2018xan,
    author = "Shadmi, Yael and Weiss, Yaniv",
    title = "{Effective Field Theory Amplitudes the On-Shell Way: Scalar and Vector Couplings to Gluons}",
    eprint = "1809.09644",
    archivePrefix = "arXiv",
    primaryClass = "hep-ph",
    doi = "10.1007/JHEP02(2019)165",
    journal = "JHEP",
    volume = "02",
    pages = "165",
    year = "2019"
}

@article{Wu:2021nmq,
    author = "Wu, Chao and Zhu, Shou-Hua",
    title = "{Massive on-shell recursion relations for n-point amplitudes}",
    eprint = "2112.12312",
    archivePrefix = "arXiv",
    primaryClass = "hep-th",
    doi = "10.1007/JHEP06(2022)117",
    journal = "JHEP",
    volume = "06",
    pages = "117",
    year = "2022"
}

@article{Alwall_2011,
   title={MadGraph 5: going beyond},
   volume={2011},
   ISSN={1029-8479},
   url={http://dx.doi.org/10.1007/JHEP06(2011)128},
   DOI={10.1007/jhep06(2011)128},
   number={6},
   journal={Journal of High Energy Physics},
   publisher={Springer Science and Business Media LLC},
   author={Alwall, Johan and Herquet, Michel and Maltoni, Fabio and Mattelaer, Olivier and Stelzer, Tim},
   year={2011},
   month=jun }

@article{Alwall_2014,
   title={The automated computation of tree-level and next-to-leading order differential cross sections, and their matching to parton shower simulations},
   volume={2014},
   ISSN={1029-8479},
   url={http://dx.doi.org/10.1007/JHEP07(2014)079},
   DOI={10.1007/jhep07(2014)079},
   number={7},
   journal={Journal of High Energy Physics},
   publisher={Springer Science and Business Media LLC},
   author={Alwall, J. and Frederix, R. and Frixione, S. and Hirschi, V. and Maltoni, F. and Mattelaer, O. and Shao, H.-S. and Stelzer, T. and Torrielli, P. and Zaro, M.},
   year={2014},
   month=jul }

@article{Actis_2013,
   title={Recursive generation of one-loop amplitudes in the Standard Model},
   volume={2013},
   ISSN={1029-8479},
   url={http://dx.doi.org/10.1007/JHEP04(2013)037},
   DOI={10.1007/jhep04(2013)037},
   number={4},
   journal={Journal of High Energy Physics},
   publisher={Springer Science and Business Media LLC},
   author={Actis, S. and Denner, A. and Hofer, L. and Scharf, A. and Uccirati, S.},
   year={2013},
   month=apr }

@article{Actis_2017,
   title={R E C O L A—REcursive Computation of One-Loop Amplitudes},
   volume={214},
   ISSN={0010-4655},
   url={http://dx.doi.org/10.1016/j.cpc.2017.01.004},
   DOI={10.1016/j.cpc.2017.01.004},
   journal={Computer Physics Communications},
   publisher={Elsevier BV},
   author={Actis, Stefano and Denner, Ansgar and Hofer, Lars and Lang, Jean-Nicolas and Scharf, Andreas and Uccirati, Sandro},
   year={2017},
   month=may, pages={140–173} }

@article{Cullen_2012,
   title={Automated one-loop calculations with GoSam},
   volume={72},
   ISSN={1434-6052},
   url={http://dx.doi.org/10.1140/epjc/s10052-012-1889-1},
   DOI={10.1140/epjc/s10052-012-1889-1},
   number={3},
   journal={The European Physical Journal C},
   publisher={Springer Science and Business Media LLC},
   author={Cullen, Gavin and Greiner, Nicolas and Heinrich, Gudrun and Luisoni, Gionata and Mastrolia, Pierpaolo and Ossola, Giovanni and Reiter, Thomas and Tramontano, Francesco},
   year={2012},
   month=mar }

@article{Cullen_2014,
   title={GoSam-2.0: a tool for automated one-loop calculations within the Standard Model and beyond},
   volume={74},
   ISSN={1434-6052},
   url={http://dx.doi.org/10.1140/epjc/s10052-014-3001-5},
   DOI={10.1140/epjc/s10052-014-3001-5},
   number={8},
   journal={The European Physical Journal C},
   publisher={Springer Science and Business Media LLC},
   author={Cullen, Gavin and van Deurzen, Hans and Greiner, Nicolas and Heinrich, Gudrun and Luisoni, Gionata and Mastrolia, Pierpaolo and Mirabella, Edoardo and Ossola, Giovanni and Peraro, Tiziano and Schlenk, Johannes and von Soden-Fraunhofen, Johann Felix and Tramontano, Francesco},
   year={2014},
   month=aug }

@article{Cascioli_2012,
   title={Scattering Amplitudes with Open Loops},
   volume={108},
   ISSN={1079-7114},
   url={http://dx.doi.org/10.1103/PhysRevLett.108.111601},
   DOI={10.1103/physrevlett.108.111601},
   number={11},
   journal={Physical Review Letters},
   publisher={American Physical Society (APS)},
   author={Cascioli, F. and Maierhöfer, P. and Pozzorini, S.},
   year={2012},
   month=mar }

@article{Buccioni_2018,
   title={On-the-fly reduction of open loops},
   volume={78},
   ISSN={1434-6052},
   url={http://dx.doi.org/10.1140/epjc/s10052-018-5562-1},
   DOI={10.1140/epjc/s10052-018-5562-1},
   number={1},
   journal={The European Physical Journal C},
   publisher={Springer Science and Business Media LLC},
   author={Buccioni, Federico and Pozzorini, Stefano and Zoller, Max},
   year={2018},
   month=jan }

@article{OL2015,
    author = {Kallweit, Stefan and Lindert, Jonas M. and Maierh{\"o}fer, Philipp and Pozzorini, Stefano and Sch{\"o}nherr, Marek},
    title = "{NLO electroweak automation and precise predictions for W+multijet production at the LHC}",
    eprint = "1412.5157",
    archivePrefix = "arXiv",
    primaryClass = "hep-ph",
    reportNumber = "LPN14-127, IPPP-14-107, DCPT-14-214, MCNET-14-26, ZU-TH-42-14, MITP-14-102",
    doi = "10.1007/JHEP04(2015)012",
    journal = "JHEP",
    volume = "04",
    pages = "012",
    year = "2015"
}

@article{Berger_2008,
   title={One-Loop Calculations with BlackHat},
   volume={183},
   ISSN={0920-5632},
   url={http://dx.doi.org/10.1016/j.nuclphysbps.2008.09.123},
   DOI={10.1016/j.nuclphysbps.2008.09.123},
   journal={Nuclear Physics B - Proceedings Supplements},
   publisher={Elsevier BV},
   author={Berger, C.F. and Bern, Z. and Dixon, L.J. and Febres Cordero, F. and Forde, D. and Ita, H. and Kosower, D.A. and Maître, D.},
   year={2008},
   month=oct, pages={313–319} }

@article{Ellis_2012,
   title={One-loop calculations in quantum field theory: From Feynman diagrams to unitarity cuts},
   volume={518},
   ISSN={0370-1573},
   url={http://dx.doi.org/10.1016/j.physrep.2012.01.008},
   DOI={10.1016/j.physrep.2012.01.008},
   number={4–5},
   journal={Physics Reports},
   publisher={Elsevier BV},
   author={Ellis, R. Keith and Kunszt, Zoltan and Melnikov, Kirill and Zanderighi, Giulia},
   year={2012},
   month=sep, pages={141–250} }

@article{Maitre_2007,
    author = "Maitre, D. and Mastrolia, P.",
    title = "{S@M, a Mathematica Implementation of the Spinor-Helicity Formalism}",
    eprint = "0710.5559",
    archivePrefix = "arXiv",
    primaryClass = "hep-ph",
    reportNumber = "SLAC-PUB-12867, ZU-TH-25-07",
    doi = "10.1016/j.cpc.2008.05.002",
    journal = "Comput. Phys. Commun.",
    volume = "179",
    pages = "501--574",
    year = "2008"
}

@article{Peraro_2014,
   title={Ninja: Automated integrand reduction via Laurent expansion for one-loop amplitudes},
   volume={185},
   ISSN={0010-4655},
   url={http://dx.doi.org/10.1016/j.cpc.2014.06.017},
   DOI={10.1016/j.cpc.2014.06.017},
   number={10},
   journal={Computer Physics Communications},
   publisher={Elsevier BV},
   author={Peraro, Tiziano},
   year={2014},
   month=oct, pages={2771–2797} }

@article{Mastrolia_2010,
   title={Scattering amplitudes from unitarity-based reduction algorithm at the integrand-level},
   volume={2010},
   ISSN={1029-8479},
   url={http://dx.doi.org/10.1007/JHEP08(2010)080},
   DOI={10.1007/jhep08(2010)080},
   number={8},
   journal={Journal of High Energy Physics},
   publisher={Springer Science and Business Media LLC},
   author={Mastrolia, P. and Ossola, G. and Reiter, T. and Tramontano, F.},
   year={2010},
   month=aug }

@article{Borowka_2018,
   title={pySecDec: A toolbox for the numerical evaluation of multi-scale integrals},
   volume={222},
   ISSN={0010-4655},
   url={http://dx.doi.org/10.1016/j.cpc.2017.09.015},
   DOI={10.1016/j.cpc.2017.09.015},
   journal={Computer Physics Communications},
   publisher={Elsevier BV},
   author={Borowka, S. and Heinrich, G. and Jahn, S. and Jones, S.P. and Kerner, M. and Schlenk, J. and Zirke, T.},
   year={2018},
   month=jan, pages={313–326} }

@article{Forde_2007,
   title={Direct extraction of one-loop integral coefficients},
   volume={75},
   ISSN={1550-2368},
   url={http://dx.doi.org/10.1103/PhysRevD.75.125019},
   DOI={10.1103/physrevd.75.125019},
   number={12},
   journal={Physical Review D},
   publisher={American Physical Society (APS)},
   author={Forde, Darren},
   year={2007},
   month=jun }

@article{Bern_1994_1,
   title={One-loop n-point gauge theory amplitudes, unitarity and collinear limits},
   volume={425},
   ISSN={0550-3213},
   url={http://dx.doi.org/10.1016/0550-3213(94)90179-1},
   DOI={10.1016/0550-3213(94)90179-1},
   number={1–2},
   journal={Nuclear Physics B},
   publisher={Elsevier BV},
   author={Bern, Zvi and Dixon, Lance and Dunbar, David C. and Kosower, David A.},
   year={1994},
   month=aug, pages={217–260} }

@article{Bern_2011,
   title={Basics of generalized unitarity},
   volume={44},
   ISSN={1751-8121},
   url={http://dx.doi.org/10.1088/1751-8113/44/45/454003},
   DOI={10.1088/1751-8113/44/45/454003},
   number={45},
   journal={Journal of Physics A: Mathematical and Theoretical},
   publisher={IOP Publishing},
   author={Bern, Zvi and Huang, Yu-tin},
   year={2011},
   month=oct, pages={454003} }

@article{Badger_2009,
   title={Direct extraction of one loop rational terms},
   volume={2009},
   ISSN={1029-8479},
   url={http://dx.doi.org/10.1088/1126-6708/2009/01/049},
   DOI={10.1088/1126-6708/2009/01/049},
   number={01},
   journal={Journal of High Energy Physics},
   publisher={Springer Science and Business Media LLC},
   author={Badger, S.D},
   year={2009},
   month=jan, pages={049–049} }

@article{Badger_2008,
   title={Generalised Unitarity at One-loop with Massive Fermions},
   volume={183},
   ISSN={0920-5632},
   url={http://dx.doi.org/10.1016/j.nuclphysbps.2008.09.107},
   DOI={10.1016/j.nuclphysbps.2008.09.107},
   journal={Nuclear Physics B - Proceedings Supplements},
   publisher={Elsevier BV},
   author={Badger, S.D.},
   year={2008},
   month=oct, pages={220–225} }

@article{Ellis_2008,
   title={A numerical unitarity formalism for evaluating one-loop amplitudes},
   volume={2008},
   ISSN={1029-8479},
   url={http://dx.doi.org/10.1088/1126-6708/2008/03/003},
   DOI={10.1088/1126-6708/2008/03/003},
   number={03},
   journal={Journal of High Energy Physics},
   publisher={Springer Science and Business Media LLC},
   author={Ellis, R.K and Giele, W.T and Kunszt, Z},
   year={2008},
   month=mar, pages={003–003} }

@article{Giele_2008,
   title={On the numerical evaluation of one-loop amplitudes: the gluonic case},
   volume={2008},
   ISSN={1029-8479},
   url={http://dx.doi.org/10.1088/1126-6708/2008/06/038},
   DOI={10.1088/1126-6708/2008/06/038},
   number={06},
   journal={Journal of High Energy Physics},
   publisher={Springer Science and Business Media LLC},
   author={Giele, W.T and Zanderighi, G},
   year={2008},
   month=jun, pages={038–038} }

@article{Hameren_2009,
   title={Automated one-loop calculations: a proof of concept},
   volume={2009},
   ISSN={1029-8479},
   url={http://dx.doi.org/10.1088/1126-6708/2009/09/106},
   DOI={10.1088/1126-6708/2009/09/106},
   number={09},
   journal={Journal of High Energy Physics},
   publisher={Springer Science and Business Media LLC},
   author={Hameren, A. van and Papadopoulos, C.G and Pittau, R},
   year={2009},
   month=sep, pages={106–106} }

@article{Bevilacqua_2009,
   title={Assault on the NLO wishlist: pp→t$\Bar{t}$b$\Bar{b}$},
   volume={2009},
   ISSN={1029-8479},
   url={http://dx.doi.org/10.1088/1126-6708/2009/09/109},
   DOI={10.1088/1126-6708/2009/09/109},
   number={09},
   journal={Journal of High Energy Physics},
   publisher={Springer Science and Business Media LLC},
   author={Bevilacqua, G and Czakon, M and Papadopoulos, C.G and Pittau, R and Worek, M},
   year={2009},
   month=sep, pages={109–109} }

@article{Bevilacqua_2013,
   title={HELAC-NLO},
   volume={184},
   ISSN={0010-4655},
   url={http://dx.doi.org/10.1016/j.cpc.2012.10.033},
   DOI={10.1016/j.cpc.2012.10.033},
   number={3},
   journal={Computer Physics Communications},
   publisher={Elsevier BV},
   author={Bevilacqua, G. and Czakon, M. and Garzelli, M.V. and van Hameren, A. and Kardos, A. and Papadopoulos, C.G. and Pittau, R. and Worek, M.},
   year={2013},
   month=mar, pages={986–997} }

@article{Binoth_2008uq,
    author = "Binoth, T. and Guillet, J. -Ph. and Heinrich, G. and Pilon, E. and Reiter, T.",
    title = "{Golem95: A Numerical program to calculate one-loop tensor integrals with up to six external legs}",
    eprint = "0810.0992",
    archivePrefix = "arXiv",
    primaryClass = "hep-ph",
    reportNumber = "EDINBURGH-2008-40, LAPTH-1277-08, IPPP-08-73, DCPT-08-146, NIKHEF-2008-26",
    doi = "10.1016/j.cpc.2009.06.024",
    journal = "Comput. Phys. Commun.",
    volume = "180",
    pages = "2317--2330",
    year = "2009"
}

@article{Giele_2009ui,
    author = "Giele, Walter and Kunszt, Zoltan and Winter, Jan",
    title = "{Efficient Color-Dressed Calculation of Virtual Corrections}",
    eprint = "0911.1962",
    archivePrefix = "arXiv",
    primaryClass = "hep-ph",
    reportNumber = "FERMILAB-PUB-09-406-T",
    doi = "10.1016/j.nuclphysb.2010.07.007",
    journal = "Nucl. Phys. B",
    volume = "840",
    pages = "214--270",
    year = "2010"
}

@article{Hahn_2000,
   title={Automatic loop calculations with FeynArts, FormCalc, and LoopTools},
   volume={89},
   ISSN={0920-5632},
   url={http://dx.doi.org/10.1016/S0920-5632(00)00848-3},
   DOI={10.1016/s0920-5632(00)00848-3},
   number={1–3},
   journal={Nuclear Physics B - Proceedings Supplements},
   publisher={Elsevier BV},
   author={Hahn, Thomas},
   year={2000},
   month=oct, pages={231–236} }

@article{Giele_2008i,
   title={Full one-loop amplitudes from tree amplitudes},
   volume={2008},
   ISSN={1029-8479},
   url={http://dx.doi.org/10.1088/1126-6708/2008/04/049},
   DOI={10.1088/1126-6708/2008/04/049},
   number={04},
   journal={Journal of High Energy Physics},
   publisher={Springer Science and Business Media LLC},
   author={Giele, Walter T and Kunszt, Zoltan and Melnikov, Kirill},
   year={2008},
   month=apr, pages={049–049} }

@article{Abreu:2021asb,
    author = "Abreu, S. and Febres Cordero, F. and Ita, H. and Klinkert, M. and Page, B. and Sotnikov, V.",
    title = "{Leading-color two-loop amplitudes for four partons and a W boson in QCD}",
    eprint = "2110.07541",
    archivePrefix = "arXiv",
    primaryClass = "hep-ph",
    reportNumber = "CERN-TH-2021-156, FR-PHENO-2021-12, MPP-2021-181",
    doi = "10.1007/JHEP04(2022)042",
    journal = "JHEP",
    volume = "04",
    pages = "042",
    year = "2022"
}

@article{DeLaurentis:2025dxw,
    author = "De Laurentis, Giuseppe and Ita, Harald and Page, Ben and Sotnikov, Vasily",
    title = "{Compact two-loop QCD corrections for Vjj production in proton collisions}",
    eprint = "2503.10595",
    archivePrefix = "arXiv",
    primaryClass = "hep-ph",
    reportNumber = "PSI-PR-24-29, ZU-TH 14/25",
    doi = "10.1007/JHEP06(2025)093",
    journal = "JHEP",
    volume = "06",
    pages = "093",
    year = "2025"
}

@article{Chicherin:2021dyp,
    author = "Chicherin, Dmitry and Sotnikov, Vasily and Zoia, Simone",
    title = "{Pentagon functions for one-mass planar scattering amplitudes}",
    eprint = "2110.10111",
    archivePrefix = "arXiv",
    primaryClass = "hep-ph",
    reportNumber = "LAPTH-041/21, MPP-2021-182",
    doi = "10.1007/JHEP01(2022)096",
    journal = "JHEP",
    volume = "01",
    pages = "096",
    year = "2022"
}

@article{Chen:2026jxf,
    author = "Chen, Xuan and Chicherin, Dmitry and Fox, Elliot and Glover, Nigel and Marcoli, Matteo and Sotnikov, Vasily and Sun, Huiting and Zhang, Hantian and Zoia, Simone",
    title = "{The Four-Jet Rate in Electron-Positron Annihilation at Order $\alpha_s^4$}",
    eprint = "2602.18185",
    archivePrefix = "arXiv",
    primaryClass = "hep-ph",
    reportNumber = "CERN-TH-2026-017, IPPP/26/12, ZU-TH 04/26, LAPTH-007/26",
    month = "2",
    journal = "",
    year = "2026"
}

@software{giuseppe_de_laurentis_2026_18895380,
  author       = {Giuseppe De Laurentis},
  title        = {GDeLaurentis/antares-results: v0.1.1},
  month        = mar,
  year         = 2026,
  publisher    = {Zenodo},
  version      = {v0.1.1},
  doi          = {10.5281/zenodo.18895380},
  url          = {https://doi.org/10.5281/zenodo.18895380},
}

\end{document}